\documentclass[pdflatex,sn-nature]{sn-jnl}
\usepackage{graphicx}%
\usepackage{multirow}%
\usepackage{amsmath,amssymb,amsfonts}%
\usepackage{amsthm}%
\usepackage[title]{appendix}%
\usepackage{xcolor}%
\usepackage{textcomp}%
\usepackage{manyfoot}%
\usepackage{booktabs}%
\usepackage{algorithm}%
\usepackage{algorithmicx}%
\usepackage{algpseudocode}%
\usepackage{listings,url}%
\usepackage{bookmark}
\usepackage{tikz}
\usepackage{xcolor}

\definecolor{stepPink}{RGB}{199, 21, 133} 

\newcommand{\stepcirc}[1]{%
    \tikz[baseline=(char.base)]{%
        \node[shape=circle, fill=stepPink, text=white, font=\bfseries, inner sep=0.4pt, minimum size=0.2em] (char) {#1};%
    }%
}

\theoremstyle{thmstyleone}%
\theoremstyle{thmstyletwo}%

\theoremstyle{thmstylethree}%

\begin{document}

\title[Article Title]{LLM-assisted Agentic Edge Intelligence Framework}

\author*[1]{\fnm{Chinmaya~Kumar} \sur{Dehury}}\email{dehury@iiserbpr.ac.in}

\author[1]{\fnm{Siddharth~Singh} \sur{Kushwaha}}\email{2510601@iiserbpr.ac.in}

\author[2]{\fnm{Qiyang} \sur{Zhang}}\email{qiyangzhang@pku.edu.cn}

\author[3]{\fnm{Alaa} \sur{Saleh}}\email{alaasa@ieee.org}

\author*[4]{\fnm{Praveen~Kumar} \sur{Donta}}\email{praveen@dsv.su.se}

\affil[1]{\orgdiv{Distributed Computing Continuum (DCC) Lab, Department
of Computer Science}, \orgname{Indian Institute of Science Education and Research Berhampur}, \orgaddress{\city{City}, \postcode{100190}, \state{State}, \country{India}}}

\affil[2]{\orgdiv{Computer Science School}, \orgname{Peking University}, \orgaddress{\city{Beijing}, \postcode{100876}, \state{State}, \country{China}}}

\affil[3]{\orgdiv{Center for Applied Computing}, \orgname{University of Oulu}, \orgaddress{\city{Oulu}, \postcode{90014}, \state{State}, \country{Finland}}}

\affil[4]{\orgdiv{Department of Computer and Systems Sciences}, \orgname{Stockholm University}, \orgaddress{\city{Stockholm}, \postcode{106 91}, \state{Stockholm}, \country{Sweden}}}


\abstract{Edge intelligence delivers low-latency inference, yet most edge analytics remain hard-coded and must be redeployed as conditions change. When data patterns shift or new questions arise, engineers often need to write new scripts and push updates to devices, which slows iteration and raises operating costs. This limited adaptability reduces scalability and autonomy in large, heterogeneous, and resource-constrained edge deployments, and it increases reliance on human oversight. Meanwhile, large language models (LLMs) can interpret instructions and generate code, but their compute and memory requirements typically prevent direct deployment on edge devices. We address this gap with the LLM-assisted Edge Intelligence (LEI) framework, which removes the need for manually specified business logic. In LEI, a cloud-hosted LLM coordinates the creation and update of device-side logic as requirements evolve. The system generates candidate lightweight programs, checks them against available data and constraints, and then deploys the selected version to each device. This lets each device receive a tailored program based on sample data, metadata, context, and current resource limits. We evaluate LEI on four heterogeneous datasets, including air quality, temperature \& humidity, wind, and soil datasets using multiple LLM backends. The experimental results show that the framework maintains low average CPU and memory utilization during the execution. These results indicate that the framework adapts efficiently to changing conditions while maintaining resource efficiency.
}

\keywords{Edge Intelligence, Large Language Models, Agentic AI, Resource Management, Context-aware Orchestration}

\maketitle

\section{Introduction}
Edge computing refers to the ability to process and analyze locally generated data without requiring constant communication with centralized cloud services~\cite{8746691}. Beyond data processing, the integration of AI with edge computing, referred to as Edge Intelligence, enables on-device decision-making, reduces latency and cloud dependency~\cite{8736011,10.1145/3724420}. For example, in smart traffic management systems, edge devices—such as smart traffic lights, cameras, and sensors—work independently to adjust traffic signals in real time based on local data about the crowd and other emergency situations. These systems can detect traffic congestion, accidents, or pedestrian movement and make decisions instantly, improving traffic flow, reducing delays, and enhancing safety~\cite{10133894}. 

Embedded analytics on edge devices support sensing and basic decision-making, but the underlying logic is usually fixed at deployment. Devices therefore rely on predefined rules and models, with limited ability to adjust when conditions shift. Changes in data distributions, sensing modalities, operational goals, or available compute and energy can quickly invalidate existing pipelines. When data formats or semantics evolve, or when analytical goals change, updates are often required not only to the processing logic but also to task scheduling and execution on-device \cite{11130884}. Hardware diversity adds another layer of difficulty, since the same workload can behave very differently across devices and may require device-specific tuning to avoid overload or failure. Similarly, introducing new sensors disrupts the entire development-to-deployment pipeline. As a result, developers must manually analyze metadata, interpret context, redesign algorithms, ensure hardware compatibility, optimize for constrained resources, and redeploy updated code to each device. This process is time-consuming, error-prone, and may require temporary service interruption. Continuous human supervision throughout the life-cycle limits scalability, delays adaptation, and increases maintenance overhead across distributed edge infrastructures~\cite{9319337}.

\begin{figure}[t]
    \centering
    \includegraphics[width=0.85\linewidth]{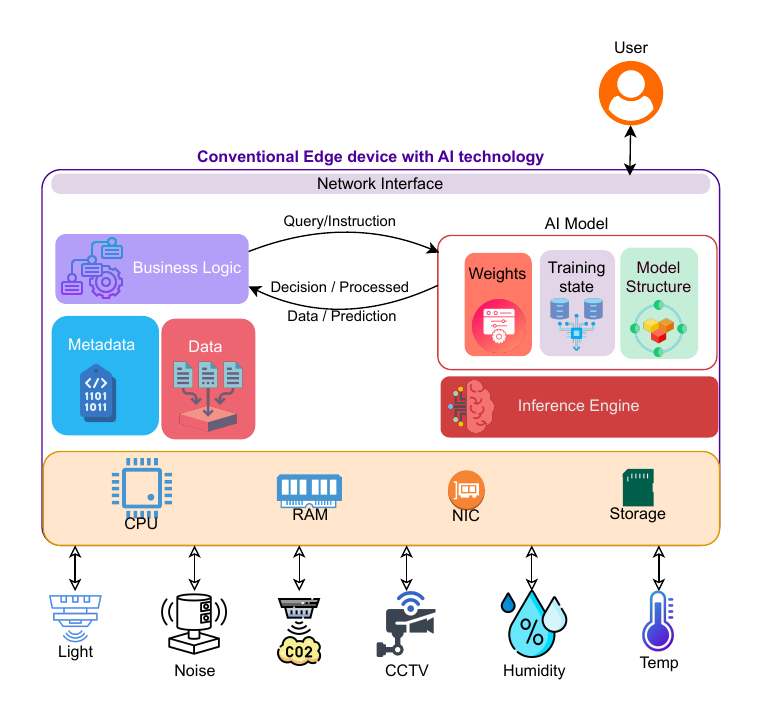}
    \caption{Conventional Edge devices with AI model}
    \label{fig:bussines}
\end{figure}
On the other hand, Large Language Models (LLMs), such as OpenAI's GPT-5.2\footnote{https://openai.com/index/chatgpt/}, Anthropic's Claude\footnote{https://claude.ai/}, or DeepSeek-V3.1\footnote{https://www.deepseek.com/en},  are growing rapidly and capable of understanding and generating human-like text based on large-scale datasets. Such models are used in a wide variety of applications ranging from natural language understanding and translation to content generation and code generation~\cite{10.1145/3635059.3635104}. These capabilities have been recently adapted in several fields, including healthcare, education, customer service, and finance, by automating tasks that require sophisticated understanding and reasoning~\cite{raza2025industrial}.  However, deploying LLMs directly (refer to Figure \ref{fig:bussines}) on edge devices is not feasible due to the computational demands of these models. Edge devices are typically resource-constrained, with limited processing power, memory, and storage, and hence, are not a suitable environment to run such large-scale models \cite{pujol2023edge}. Edge device hardware and software stacks generally cannot accommodate models of that size, even when real-time inference is not a requirement~\cite{10.1145/3788870}.

Given this, LLMs can be leveraged in edge computing systems through two primary strategies: (i) compressing and optimizing models for direct on-device deployment, and (ii) enhancing edge hardware to satisfy LLM resource requirements. In the first approach, techniques such as quantization, pruning, and distillation reduce memory footprint and computational demand, making inference feasible on resource-constrained devices; however, aggressive compression often degrades accuracy, robustness, and reasoning performance by reducing representational capacity and limiting effectiveness on complex or out-of-distribution tasks. In the second approach, equipping edge nodes with more capable processors (e.g., GPUs/NPUs), larger memory, and greater storage can sustain higher-quality LLM inference, but this increases capital and operating costs, energy consumption, and thermal constraints, and it does not scale easily across large, heterogeneous edge infrastructures—making widespread deployment economically impractical in many settings~\cite{10.1145/3788870}. While both approaches have their merits, they also present significant challenges that must be carefully addressed.  
In aggregate, these limitations motivate a shift from purely model- or hardware-centric solutions toward agentic, LLM-assisted orchestration frameworks that enable edge systems to adapt, optimize, and evolve autonomously under tight resource constraints. 

There is increasing demand for a framework that enables edge devices to autonomously upgrade their intelligence using LLMs, with minimal to no human involvement. Such a framework should close the gap between static, hard-coded edge logic and adaptive, continuously improving edge behavior. In response, we propose \textit{LLM-assisted Edge Intelligence} (LEI) — an agentic architecture in which the LLM serves as a cognitive orchestrator. LEI ingests sample data, metadata, and contextual signals; profiles device-side constraints (e.g., compute, memory, energy, and latency); synthesizes resource-aware, task-specific programs; and deploys them back to the edge for local execution. This design supports continuous capability evolution while preserving low latency and efficient resource use. Agentic decision-making at the orchestration layer further allows edge nodes to respond to environmental changes and workload shifts without slowing time-critical inference. Our contributions are summarized as follows:
\begin{enumerate}
    \item We have designed and developed a framework that dynamically generates business logic based on real-time context using LLMs, minimizing the need for manual intervention.
    \item To enhance system robustness and security, a dedicated component is introduced to validate faulty business logic, developed by LLMs, before further processing.
    \item Our proposed framework is adaptive in nature, which generates lightweight code and consumes minimal CPU and memory usage based on the available resources of edge devices, allowing users to perform other tasks without worrying about the resources of edge devices. 
\end{enumerate}

The remainder of this paper is organized as follows: Section ~\ref{sec:rel_work} reviews the existing literature on large language models (LLMs) and their deployment on resource-constrained devices. Section~\ref{sec:syArch} presents a high-level overview of the proposed architecture. Section~\ref{sec:exper} describe the implementation across  different use cases, models and performance metrics. The experimental results are presented in Section~\ref{sec:res}. Finally, Section~\ref{sec:con} concludes the paper. Table-\ref{tab:acronym} provides the list of acronyms used in this paper.

\begin{table}[htb]\label{tab:acronym}
\caption{List of Acronyms}
\begin{tabular}{|@{}l|l|}
\hline
\textbf{Acronym} & \textbf{Description} \\ \hline \hline
API	& Application Programming Interface \\ \hline
AQ	 & Air Quality \\ \hline
BERT 	& Bidirectional Encoder Representations from Transformers\\ \hline
CPU	& Central Processing Unit \\ \hline
CSV	&  Comma Separated Values\\ \hline
GPU	& Graphics Processing Unit \\ \hline
I/O	& Input and Output\\ \hline
JSON &  JavaScript Object Notation\\ \hline
LEI	& LMM-assisted Edge Intelligence \\ \hline
LLM	& Large Language Model \\ \hline
NPU	& Neural Processing Unit \\ \hline
OS	 & Operating Systems \\ \hline
RAM	& Random Access Memory \\ \hline
TH	 & Temperature and Humidity \\ \hline
\end{tabular}
\end{table}

\section{Related Works}\label{sec:rel_work}

As LLM capabilities improve and edge deployments expand, researchers have increasingly explored executing LLM workloads directly on resource-limited devices~\cite{10.1145/3719664,10.1145/3736721}. On one hand, collaborative intelligence among heterogeneous agents uses dynamic orchestration frameworks to enable effective multi-agent coordination, achieving task performance that surpasses single-model systems \cite{jain2020spatula, ma2024hpipe}. However, deploying such collaborative frameworks on edge devices remains challenging due to stringent constraints on communication bandwidth, latency, and energy efficiency \cite{lu2023multi, tian2024large}. Existing collaborative frameworks typically rely on manually designed business logic or pre-defined analytical pipelines, which limits their adaptability to dynamic environments and heterogeneous edge devices~\cite{10.1145/3719664}.

On the other hand, efficiently deploying and scheduling for single-device inference adopts a layered strategy, where multiple techniques collaboratively address optimization challenges from complementary perspectives. Recent advancements can be broadly categorized into token reduction, early exiting, and dynamic offloading. Token reduction aims to decrease the computational overhead by selectively eliminating redundant or less informative tokens from the inputs. For instance, PoWER-BERT \cite{51goyal2020power} removes non-critical word representations, while Length-Adaptive Transformer \cite{88kim2023language} dynamically adjusts input sequence lengths through a dropout mechanism. In addition, LLMLingua \cite{79jiang2023llmlingua} performs iterative prompt compression at the token level, and AutoCompressors \cite{29chevalier2023adapting} condense long context windows into compact summary vectors. These methods reduce input size, improve memory efficiency, and alleviate computational burden, which is particularly beneficial for resource-constrained edge devices.

Early exiting reduces inference latency by terminating computation once a predefined confidence threshold is reached, thereby avoiding unnecessary forward-pass operations. This strategy becomes even more effective when combined with reduced input sizes.
For example, PABEE \cite{231zhou2020bert} improves both efficiency and robustness by integrating internal classifiers at each transformer layer. FREE \cite{11bae2023fast} introduces a shallow-deep module that synchronizes the decoding of the current token with previously exited tokens, while ConsistentEE \cite{216zeng2024consistentee} employs reinforcement learning to identify optimal exit points, achieving a balance between inference speed and prediction accuracy.

Dynamic offloading further enhances resource utilization by distributing inference workloads across heterogeneous computing resources. 
STI \cite{59guo2023sti} proposes elastic pipelining, dynamically sharding model components and allocating them based on available resources. FlexGen \cite{154sheng2023flexgen} extends this approach with a hybrid memory model that utilizes GPU, CPU, and disk memory, supported by intelligent I/O scheduling to offload computationally intensive tasks. Together, these techniques balance system workloads and enable efficient large-scale LLM inference on edge devices while minimizing performance bottlenecks.

Although recent studies have explored automated model selection or lightweight model adaptation, they often require significant human intervention during deployment and reconfiguration, and lack the capability to autonomously generate task-specific intelligence at runtime. In contrast, our proposed LEI framework eliminates the need for manual business logic design by leveraging cloud-hosted LLMs to dynamically generate, validate, and deploy lightweight analytical programs on edge devices. 
\section{LEI System Architecture}\label{sec:syArch}

\begin{figure}[t]
    \centering
    \includegraphics[width=0.50\linewidth]{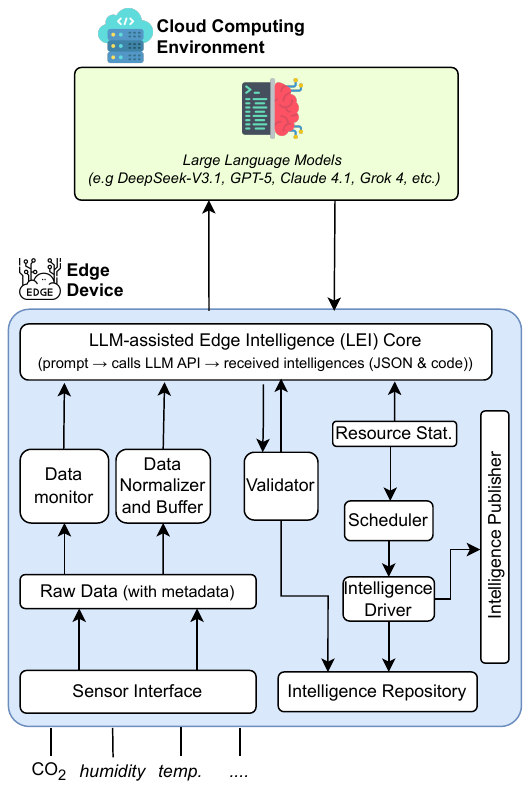}
    \caption{General system architecture of LLM-assisted Edge Intelligence (LEI) and its internal components.}
    \label{fig:architecture}
\end{figure}

\begin{figure}[t]
    \centering
    \includegraphics[width=1\textwidth]{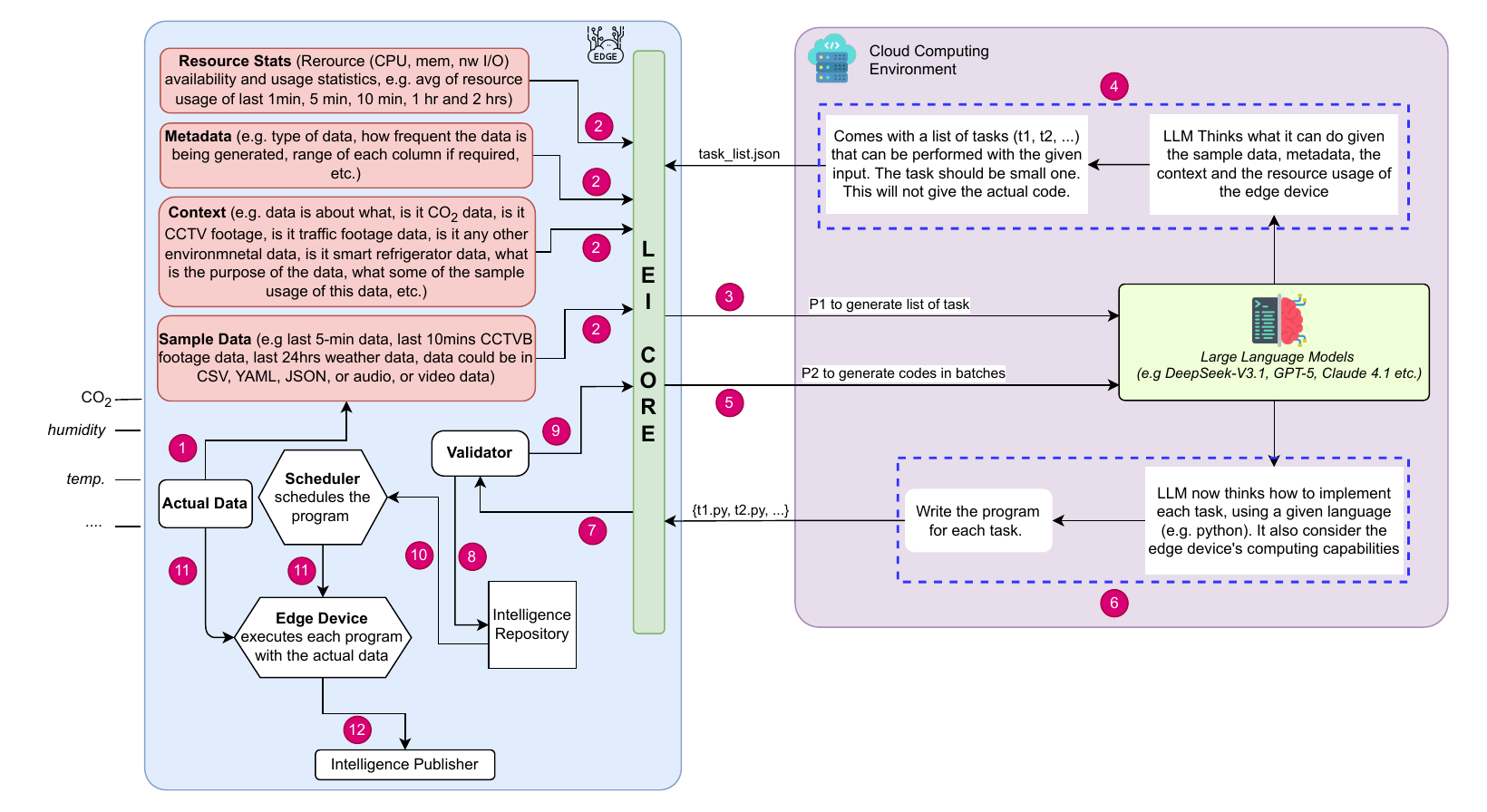}
    \caption{End-to-end workflow of LLM-assisted Edge Intelligence.}
    \label{fig:workflow}
\end{figure}

This section describes the proposed LEI framework, which uses an LLM as a cognitive orchestrator to shift edge intelligence from static, developer-defined logic to adaptive intelligence driven by real-time requirements. In LEI, application logic is not hard-coded; instead, the LLM derives task logic from representative sample data, metadata, contextual information, and device resource statistics. LEI also integrates a system health monitoring pipeline that (i) guides resource-aware task and code generation and (ii) supports scheduling decisions that mitigate edge hardware constraints. In addition, accommodating new data formats typically requires only lightweight updates to metadata and context rather than extensive code rewrites. Fig.~\ref{fig:architecture} summarizes the architecture, which comprises an input/output interface around two primary domains, namely the \textbf{Edge Execution Layer} and the \textbf{Cloud Computing Environment}. The end-to-end execution flow between these domains is illustrated in Fig.~\ref{fig:workflow}  using an environmental monitoring example.

\subsection{Edge Execution Layer}
The edge execution layer integrates the \textit{Sensor Interface}, \textit{Data Monitor}, \textit{Data Normalizer and Buffer}, \textit{Resource Stat}, \textit{LEI core}, \textit{Validator}, \textit{Scheduler}, \textit{Intelligence Driver}, \textit{Intelligence Repository}, and \textit{Intelligence Publisher}, enabling local processing under constrained CPU and memory resources.

The workflow begins at the \textit{Sensor Interface} (Fig.~\ref{fig:workflow}, \stepcirc{1}--\stepcirc{2}), which continuously collects raw streams from the environment (e.g., temperature and humidity) and stores them in formats such as CSV or JSON. The raw stream serves two roles. First, the system extracts a representative subset to form sample data (e.g., the most recent five minutes), which captures key characteristics of the full stream and supports downstream task and code synthesis. Second, the same raw stream acts as the input on which the validated code ultimately executes on-device.

After acquisition, the \textit{Data Monitor} tracks data arrival and screens for noise, while the \textit{Data Normalizer and Buffer} performs preprocessing and buffering to prepare the stream for orchestration. In parallel, the \textit{Resource Stat} component continuously monitors CPU utilization, memory usage, and device availability, and computes average CPU and memory statistics over $1$, $5$, $10$, and $30$ minute windows. These signals guide (i) the LLM in deciding how many tasks to propose or generate under device constraints and (ii) the \textit{Scheduler} in selecting feasible execution times and target devices.


\begin{algorithm}[t]
    \caption{LEI Pipeline}
    \label{alg:lei_pipeline}
    \begin{algorithmic}[1]
        \Require Sample data $S$, Metadata $D_{meta}$, Context $C_{bg}$, Prompt $P$, Resource Statistics $R_{stat}$, Previous Task $T'$, Model $M$, Edge Device $E$
        \Ensure Execution results R
        \State $T \gets$ \Call{TaskGenerator}{$M, S, D_{meta}, C_{bg}, P, R_{stat},  T'$} \label{line:ltg}
        \State $C \gets$ \Call{CodeGenerator}{$M, T, S, D_{meta}, C_{bg}, P$} \label{line:lcg}
        \State $V \gets$ \Call{Validator}{$M, C, S, P$} \label{line:lv}
        \State $R \gets \emptyset$      \label{line:m4}
        \For{each $v_i \in V$}
            \State $res \gets \Call{CheckResource}{R_{stat}}$
            \If{$res = \textsc{Sufficient}$}
                \State $r_i \gets \Call{Execute}{v_i, E}$
                \State $R \gets R \cup \{r_i\}$
            \EndIf
        \EndFor                         \label{line:m11}
        \State \Call{Send}{R, IntelligencePublisher}    \label{line:m12}
    \end{algorithmic}
\end{algorithm}

\begin{algorithm}[b]
    \caption{TaskGenerator: Generates a list of tasks based on user instructions}
    \label{alg:task_gen}
    \begin{algorithmic}[1]
        \Require Sample data $S$, Metadata $D_{meta}$, Context $C_{bg}$, Prompt $P$, Resource Statistics $R_{stat}$, Previous Task $T'$, Model $M$
        \Ensure Task List $T = \{t_1,\dots, t_n\} $
        \State $User_{instr} \gets \text{Aggregate}(S, D_{meta}, C_{bg}, P, R_{stat})$
        \State $T_{new} \gets \Call{InvokeLLM}{M, User_{instr}}$
        \State $T \gets T' \cup T_{new}$
        \State \Return $T$
    \end{algorithmic}
\end{algorithm}

\begin{algorithm}[t]
    \caption{CodeGenerator: Generate code for the task list}
    \label{alg:code_gen}
    \begin{algorithmic}[1]
        \Require Task List $T = \{t_1, \dots, t_n\}$, Sample data $S$, Metadata $D_{meta}$, Context $C_{bg}$, Prompt $P$, Model $M$
        \Ensure Code List $C = \{c_1, \dots, c_m\}$ where $m \le n$
        
        \State $C \gets \emptyset$
        
        \State Let $k$ be the defined batch size
        \State Let $b \gets \lceil n/k \rceil$ be the total number of batches
        \State Partition $T$ into $b$ subsets $B = \{B_1, \dots, B_b\}$ such that $|B_i| \le k$
        
        \For{each batch $B_i \in B$}
            \State $User_{instr} \gets \text{Aggregate}(B_i, S, P)$
            \State $c_i \gets \Call{InvokeLLM}{M, User_{instr}}$
            \State $C \gets C \cup \{c_i\}$
        \EndFor
        
        \State \Return $C$
    \end{algorithmic}
\end{algorithm}

\begin{algorithm}[b]
\caption{Validator: Validate code for the code list}
\label{alg:validator}
\begin{algorithmic}[1]
\Require Code list $C = \{c_1, \dots, c_m\}$, Sample data $S$, Model $M$
\Ensure Validated code $V = \{v_1, \dots, v_p\}$ where $p \le m$

\State $V \gets \emptyset$
\State Let $A_{max} \ge 1$ be the maximum attempts

\For{each code snippet $c_i \in C$}
    \State $status, err\_msg \gets \Call{ExecuteLocally}{c_i, S}$ 
    \If{$status = \textsc{Success}$}
        \State $V \gets V \cup \{c_i\}$
    \Else

        \For{$attempt = 1$ to $A_{max}$} \label{line:l8}       \Comment{correction loop}
            \State $User_{instr} \gets \Call{Aggregate}{c_i, err\_msg}$
            \State $c_i \gets \Call{InvokeLLM}{M, User_{instr}}$ 

            \State $status, err\_msg \gets \Call{ExecuteLocally}{c_i, S}$

            \If{$status = \textsc{Success}$}
                \State $V \gets V \cup \{c_i\}$
                \State \textbf{break}
            \EndIf
        \EndFor \label{line:l16}
    \EndIf
\EndFor

\State \Return $V$
\end{algorithmic}
\end{algorithm}

The \textit{LEI core} is the coordination hub on the edge (Algorithm~\ref{alg:lei_pipeline}). It aggregates and structures the orchestration inputs, including sample data, metadata (data type/modality, sampling frequency, sensor details, data source identifiers, and units), context (deployment environment, purpose, and required intelligence), and resource statistics (e.g., CPU load and memory utilization on a Raspberry Pi). When available, it also includes a list of previously generated tasks to enable reuse and iterative refinement; this task history may be absent on first invocation for a newly deployed sensor stream. The LEI core forwards the structured prompt to the cloud-hosted LLM, which generates task proposals (Algorithm~\ref{alg:task_gen}; Algorithm~\ref{alg:lei_pipeline}, Line~\ref{line:ltg}) and corresponding code (Algorithm~\ref{alg:code_gen}; Algorithm~\ref{alg:lei_pipeline}, Line~\ref{line:lcg}).

Once the edge device receives code, the \textit{Validator} executes it locally against the sample data to detect runtime errors or unmet requirement (Algorithm~\ref{alg:validator}; Algorithm~\ref{alg:lei_pipeline}- Line~\ref{line:lv}); (Fig.~\ref{fig:workflow}, \stepcirc{7}--\stepcirc{9}). If validation fails, the Validator returns the code and error message to the LEI core for regeneration. To prevent non-terminating correction loops, the Validator rejects code if regeneration fails beyond a bounded number of attempts (e.g., 2 in our experiment discussed in later sections) (Algorithm~\ref{alg:validator}, Lines~\ref{line:l8}--~\ref{line:l16}). Validated code is stored in the \textit{Intelligence Repository} for reuse and scheduling.

Execution is controlled by the \textit{Scheduler} (Algorithm \ref{alg:lei_pipeline}, Lines \ref{line:m4}--\ref{line:m11}), which uses \textit{Resource Stat} details to ensure that the edge device has sufficient CPU and memory headroom to avoid overload-related crashes (Fig.~\ref{fig:workflow}, \stepcirc{10}--\stepcirc{12}). The Scheduler triggers the \textit{Intelligence Driver} to execute validated code in an isolated runtime (e.g., a virtual environment) using the live raw stream. Finally, the \textit{Intelligence Publisher} collects execution outputs and exposes stakeholder-facing intelligence through dashboards or persistent storage backends (Algorithm~\ref{alg:lei_pipeline}, Line~\ref{line:m12}).

\subsection{Cloud Computing Environment}
The cloud computing environment hosts the LLM and supports edge devices by generating task plans and code artifacts that respect edge constraints. After receiving a structured prompt from the LEI core (Fig.~\ref{fig:workflow}, \stepcirc{3}--\stepcirc{4}), the LLM first analyzes the sample data, metadata, context, resource statistics, and (optionally) the task history to determine a feasible set of tasks. At this stage, the LLM proposes tasks in a structured format (e.g., \texttt{task\_list.json}) rather than producing code immediately, which limits unnecessary token usage and reduces iteration cost.
To stay within the LLM context window and maintain controllable generation, the LEI core requests code in small batches (Fig.~\ref{fig:workflow}, \stepcirc{5}--\stepcirc{6}), for example two tasks per request, using an additional prompt that specifies the desired language (e.g., Python), output structure, and exception handling behavior (including cases where no task should be generated). The LLM then produces task-specific, resource-aware implementations that avoid heavyweight dependencies and align with the target edge device budget. When the edge-side Validator reports failures, the LLM receives the corresponding error information via the LEI core and regenerates corrected code within a bounded retry budget. Overall, LEI reduces developer intervention for evolving edge deployments by centralizing task reasoning and code synthesis in the cloud-hosted LLM while enforcing correctness and feasibility through edge-side validation and scheduling.

\section{Experiments and Results}\label{sec:exper}
In this section, we present a detailed discussion of our experimental setup along with the numerical results across various performance metrics.
\subsection{Experiment setup}
In this section, we describe the experimental setup, including the hardware and software configurations, the LLMs evaluated, the use cases considered, and the datasets used for each use case.
\subsubsection{Hardware Details}
To implement our proposed LEI framework, we have utilized two separate Raspberry Pi models (Pi5 and 4B) to compare the performance under different hardware resource constraints. 
The Raspberry Pi 4B features a 1.8 GHz quad-core Cortex-A72 CPU, 1-8 GB LPDDR4 RAM, Gigabit Ethernet, dual micro-HDMI ports (up to 4Kp60), and a microSD for storage. The Raspberry Pi 5 provides more advanced/enhanced features, including a 2.4 GHz quad-core Cortex-A76 CPU, 2-16 LPDDR4X-4267 RAM, and Gigabit Ethernet with PoE+ support. Both of these single-board computers (SBCs) include the standard 40-pin GPIO, dual-band 802.11ac Wi-Fi, Bluetooth 5.0 and were connected to the internet via Ethernet LAN. 

\subsubsection{Software Details}
 In both edge devices, Debian Version 13 (Trixie) Raspberry Pi OS\footnote{https://www.raspberrypi.com/software/operating-systems/}, released on December 4, 2025, was installed. The entire LEI framework is written in Python (version 3.13.7) programming language and the Integrated Development Environment (IDE) used for development and debugging is Visual Studio Code (Version 1.106.3).

\subsubsection{Models}
To eliminate the need for users to write business logic, we have used lightweight LLMs on private cloud infrastructure. These lightweight LLM models, as shown in Table \ref{tab:ollama_models}, are hosted on the local Server with 128GB of RAM without GPU. All the deployed models are free and open-source, deployed using the OLLAMA\footnote{https://ollama.com/} framework. We have selected a set of edge-compatible models with a parameter count of less than 10 billion, capable of understanding text \& and generating code, and to ensure a balance between computational efficiency and understanding capability. Out of 8 models used, 4 models (codegemm\footnote{https://ollama.com/library/codegemma}, codellama\footnote{https://ollama.com/library/codellama}, deepseek-coder\footnote{https://ollama.com/library/deepseek-coder}, qwen2.5-coder\footnote{https://ollama.com/library/qwen2.5-coder}) can be categorized as \textit{code-based} models, as they are specifically designed for code generation and code fixing based on the instructions, while rest models (mistral\footnote{https://ollama.com/library/mistral}, gemma\footnote{https://ollama.com/library/gemma}, phi3\footnote{https://ollama.com/library/phi3}, llama3.1\footnote{https://ollama.com/library/llama3.1}) are \textit{general-purpose} models with the ability to generate and fix code as well.

\begin{table}[ht!]
    \centering
    \caption{Edge-Compatible LLMs}
    \label{tab:ollama_models}
    \renewcommand{\arraystretch}{1.2}
    \begin{tabular}{llcc} 
        \toprule
        \textbf{ID} & \textbf{Model Name \& Pramater Count} & \textbf{Size (GB)} & \textbf{context}\\
        \midrule
        M1 & codegemma:7b        & 5.0 GB & 8K \\
        M2 & codellama:7b        & 3.8 GB & 16K \\
        M3 & deepseek-coder:6.7b & 3.8 GB & 16K \\
        M4 & gemma3:4b           & 3.3 GB & 128K \\
        M5 & llama3.1:latest     & 4.9 GB & 128K \\
        M6 & mistral:7b          & 4.4 GB & 256K \\
        M7 & phi3:3.8b            & 2.2 GB & 128K \\
        M8 & qwen2.5-coder:7b    & 4.7 GB & 32K \\
        \bottomrule        
    \end{tabular}
\end{table}

\subsubsection{Data Source}
    We have utilized data from the OpenWeather\footnote{https://openweathermap.org/api} API and Open-Meteo API\footnote{https://open-meteo.com/}. OpenWeather is a widely recognized API provider to fetch weather data that offers both free and paid plans for users. On the other hand, Open-Metro offers free access to its API for non-commercial use. Table \ref{tab:data_sources} provides an overview of the data sources for each of the use cases which is discussed in section \ref{sub: use_case}.

\begin{table}[htbp]
\caption{Details of Data Sources}\label{tab:data_sources}
\begin{tabular}{@{}lll@{}}
\toprule
\textbf{Use Case} & \textbf{Location} & \textbf{Source \& Interval} \\
\midrule
Air Quality (AQ) & Delhi, India ($28.70^\circ$N, $77.10^\circ$E) & OpenWeatherMap (Real-time, 10 min) \\
Temp \& Humidity (TH) & Delhi, India ($28.70^\circ$N, $77.10^\circ$E) & OpenWeatherMap (Real-time, 10 min) \\ 
Wind (WIND) & Berhampur, India ($21.55^\circ$N, $86.64^\circ$E) & Open-Meteo (Past 16 Days, Hourly) \\ 
Soil (SOIL) & Berhampur, India ($21.55^\circ$N, $86.64^\circ$E) & Open-Meteo (Past 16 Days, Hourly) \\ 
\botrule
\end{tabular}
\end{table}

\subsection{Step-by-Step Execution Pipeline}
\begin{figure*}[htbp]
    \centering
    \includegraphics[width=0.95\textwidth]{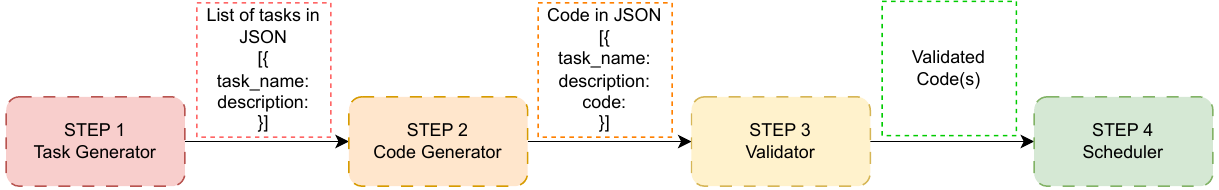}
    \caption{Step-by-Step Execution of Pipeline}
    \label{fig:sbs}
\end{figure*}

Algorithm \ref{alg:lei_pipeline} discuss the theoretically workflow of LEI framework which also can be summarized as Figure \ref{fig:sbs}. This subsection gives practical overview of proposed framework, starting from availability of entire code in the GitHub repository~\cite{github}. For easy navigation of the repository, we have summarized it in the Table~\ref{tab:repo_structure}. The LEI framework works on a general pipeline regardless of any specific data type. First, we need to set a single environment variable (e.g., \texttt{DATA\_TYPE = "air\_quality"}) in the \texttt{config.py} file. This helps the LLM to target the specific folder containing the required files (Sample Data, Metadata, and Context). \textit{resource\_usage\_summary.json} file contains the resource statistics of the edge device, which is share across the different components of the LEI framework. This file contains CPU usage (over 5-minute, 10-minute, and 30-minute periods), available memory, and the number of CPU cores on the edge device. 
\begin{table}[htbp]
\centering
\caption{Directory Structure of the Project Repository}
\begin{tabular}{@{}lp{11cm}@{}}
\toprule
\textbf{Directory} & \textbf{Description} \\
\midrule
\texttt{/data} & Contains all domain-specific datasets used in the proposed work. It includes four subfolders: air\_quality, soil, wind, and temp\_humidity. Each subfolder contains metadata, context information, and sample datasets. \\
\texttt{/prompts} & Stores all prompt used in the system, including task generation prompts, code generation prompts, and validation prompts. \\
\texttt{/output} & Stores the final outputs generated by the system after code execution. \\
\texttt{/logs} & Contains execution logs and runtime details for traceability and debugging. \\
\botrule
\end{tabular}
\label{tab:repo_structure}
\end{table}

Figure \ref{fig:sbs} shows the sequential flow of the proposed pipeline of LEI and explanation as follows:

\paragraph{Step 1: Task Generation}
The execution begins with the \textit{task\_generator.py} script. This script reads domain-specific files (sample data, metadata, context) along with resource statistics and dispatch a structured prompt to the cloud-hosted LLM to generate list of tasks as defined in Algorithm \ref{alg:task_gen}. The LLM returns a list of tasks in a JSON file named \textit{new\_tasks.json}.

\paragraph{Step 2: Code Generation}
Once all the tasks are known, the LEI framework calls the Python code-generation (see Algorithm \ref{alg:code_gen}) file \textit{code\_generator.py}. In this step, we send another structured prompt for code generation to the LLM for generating a lightweight Python script for each task. The LLM returns the task name, description, and Python code as properties of a JSON object. The \textit{code\_generator.py} normalizes the code string by removing Markdown fences and unescaping sequences, such as \textbackslash{}n to ensure a clean Python code and places it under the same environment variable folder name defined at the beginning. However, the LLM may generate incorrect code that needs to be verified before use in a real environment by the edge devices. 

\paragraph{Step 3: Validation}
Once all the code is stored, the \textit{validator} (Algorithm \ref{alg:validator}) executes the generated code locally to check for any errors. If the code fails either due to syntax errors or incorrect output formats, the validator sends the error message along with the code to the LLMs to regenerate it. To avoid getting stuck in an infinite loops caused by the LLM hallucinations (which may lead to generating wrong code again and again) or by slow internet connection, the Validator will call the LLM only twice for each task and a strict timeout timer of 120 seconds is set for each LLM call; after that, it will mark the task as failed in a separate \textit{validator\_summary.json} file. If the code is fixed after the first or second attempt, it will mark the task code as passed. After step 3, it is ensured that only working, error-free code is stored and will be used in the future for deployment.

\paragraph{Step 3: Execution \& Visualization}
The Edge schedulers, \textit{edge\_scheduler\_sequentially.py}, run the validated, error-free Python scripts sequentially on the original raw data stored in \textit{raw\_data.csv} on  available edge device as describe in Algorithm~\ref{alg:lei_pipeline}, Lines~\ref{line:m4}--\ref{line:m11}. During execution, the scheduler generates a log file (\textit{edge\_execution\_YYYYMMDD.log}) that records the start time, duration, and exit status (SUCCESS/FAILURE) of each task for each run count. The log format and examples are available in the \texttt{/log/} directory of the GitHub repository~\cite{github}. Finally, the intelligence publisher parses the latest output log and displays the content in graphical form for specific insights.

\subsection{Use case Implementation} \label{sub: use_case}
To demonstrate the working of the LEI framework, we have implemented four distinct use cases on edge devices: Air Quality, Temperature \& Humidity, Soil and Wind Monitoring. First, we will outline the common structure for all use cases, then discuss each implementation in detail. 
\subsubsection{Use case 1: Air Quality Monitoring}

\paragraph{Objective \& Data Source}
The goal of this use case is to extract insights from environmental data without manual coding. We utilized the Open Weather Map\footnote{https://openweathermap.org/api} API to collect raw data every 10 minutes for a location in Delhi (latitude $28.70^\circ$N, longitude $77.10^\circ$E). The parameters include Particulate Matter (PM$_{2.5}$, PM$_{10}$), Nitrogen Dioxide (NO$_2$), Ozone (O$_3$), Sulphur Dioxide (SO$_2$), and Ammonia (NH$_3$). Table \ref{tab:air_quality} presents a subset of the collected data.

\begin{table}[htbp]
\caption{Air Quality Sample Data}\label{tab:air_quality}

\begin{tabular}{@{}lrrrrrrrrr@{}}
\toprule
\textbf{Time} & \textbf{AQI} & \textbf{CO} & \textbf{NO} & \textbf{NO$_2$} & \textbf{O$_3$} & \textbf{SO$_2$} & \textbf{PM$_{2.5}$} & \textbf{PM$_{10}$} & \textbf{NH$_3$} \\
\midrule
2025-10-28 17:24:12 & 4 & 394.08 & 0.01 & 6.74 & 105.05 & 1.18 & 54.77 & 59.38 & 0.93 \\
2025-10-28 17:38:26 & 4 & 394.08 & 0.01 & 6.74 & 105.05 & 1.18 & 54.77 & 59.38 & 0.93 \\
2025-10-29 09:14:41 & 4 & 329.72 & 0.84 & 5.65 & 74.62  & 3.55 & 56.36 & 67.58 & 0.26 \\
2025-10-29 09:24:42 & 4 & 329.72 & 0.84 & 5.65 & 74.62  & 3.55 & 56.36 & 67.58 & 0.26 \\
2025-10-29 09:34:42 & 4 & 329.72 & 0.84 & 5.65 & 74.62  & 3.55 & 56.36 & 67.58 & 0.26 \\
2025-10-29 09:44:43 & 4 & 329.72 & 0.84 & 5.65 & 74.62  & 3.55 & 56.36 & 67.58 & 0.26 \\
\bottomrule
\end{tabular}
\end{table}

\paragraph{Domain Specific Files}
After setting the environment variable \texttt{DATA\_TYPE = "air\_quality"} in the \textit{config.py} file, we provide the following files, which are stored inside \textit{air\_quality folder}, to the LLM for a better understanding of the air quality domain: 

 \begin{itemize}
    \item \textit{sample\_data.csv}: A subset of air quality data (as shown in table \ref{tab:air_quality}.
    \item \textit{meta\_data.json} A JSON file defining units (e.g., \textmu g/m$^3$) and column descriptions. A snippet is shown in Figure \ref{fig:metadata}, while the complete metadata structure is available in Appendix \ref{app:metadata_AirQuality}.
    \item \textit{context.txt} A text file describing data collection goals, such as monitoring sudden spikes. A snippet is shown in Figure \ref{fig:context} and the complete context text in Appendix \ref{app:context_AirQuality}.
    \item \textit{tasks\_list.json} (optional) List of tasks that are already generated is stored in this JSON file, which basically helps the LLM not to generate the same or similar tasks.
\end{itemize}

\begin{figure}[htbp]
    \centering
    \includegraphics[width=0.9\linewidth]{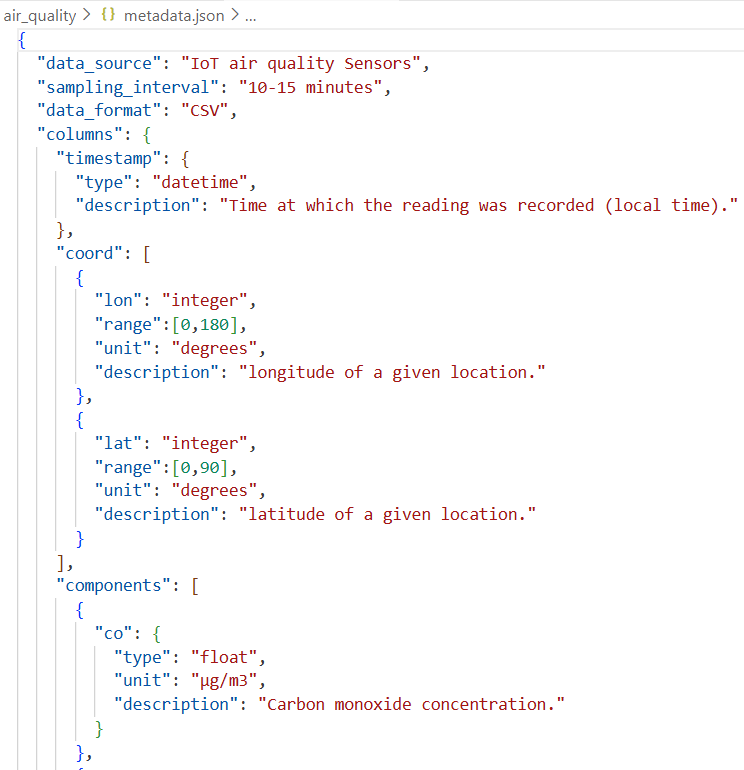}
    \caption{Metadata Air Quality in JSON format}
    \label{fig:metadata}
\end{figure}

\begin{figure}[htbp]
    \centering
    \includegraphics[width=0.9\linewidth]{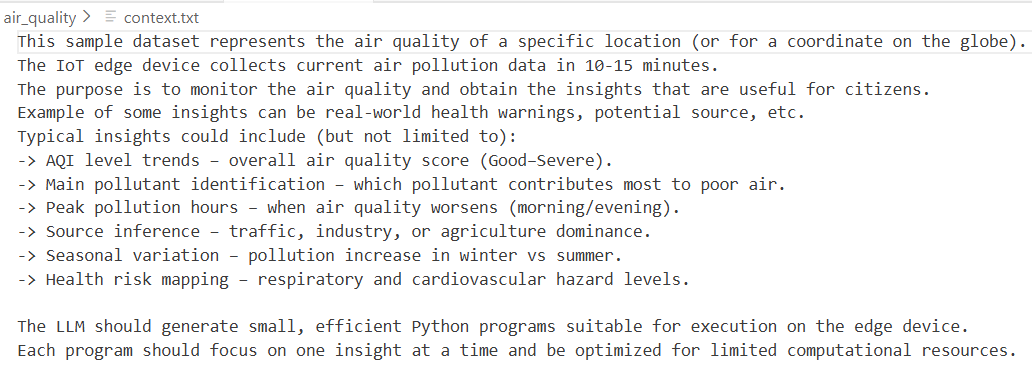}
    \caption{Context of Air Quality Dataset}
    \label{fig:context}
\end{figure}

\begin{figure}[htbp]
    \centering
    \includegraphics[width=0.5\linewidth]{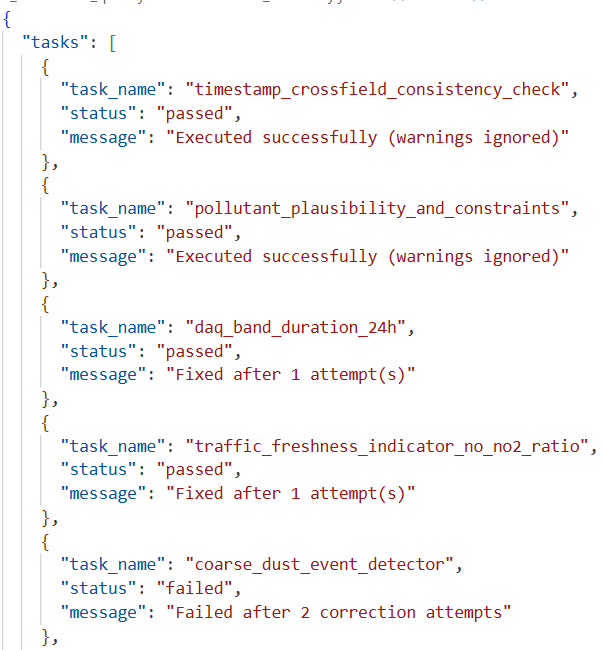}
    \caption{Validator Summary of first six tasks of air quality}
    \label{fig:val_summ}
\end{figure}

\paragraph{Pipeline Steps}
Along with the \texttt{Domain Specific Files}, the resource statistics of the edge device (\textit{resource\_usage\_summary.json}) are sent to the LLM for task generation. For example, the LLM returns tasks such as ``pollutant\_24h\_extremes\_reporter'' and ``traffic\_freshness\_indicator\_no\_no2\_ratio''. After receiving the list of task \textit{new\_task.json}, the \textit{task\_code\_generator.py} runs and generates respectively code for each task such as \textit{pollutant\_24h\_extremes\_reporter.py} and \textit{traffic\_freshness\_indicator\_no\_no2\_ratio.py}. Once the code is generated, it is validated by the \textbf{\textit{validator.py}} and generates the \textit{validator\_summary.json} file, as shown in Figure \ref{fig:val_summ}. If no error arises, then it will be stored in the intelligence repository (as shown in the Figure \ref{fig:workflow}). Finally, the Scheduler executes the validated codes, generates the log file, and sends the results for publication. 

\subsubsection{Use Case 2: Temperature \& Humidity Monitoring}

\paragraph{Objective \& Data Source}
Similar to the first use case, the objective here is to extract insights from temperature and humidity data. We used the OpenWeatherMap API to collect data for the same location, Delhi. The dataset contains three columns, namely Time, Temperature ($^\circ$C) and Humidity (\%). Table \ref{tab:temp_humidity} presents a subset of the collected data.

\begin{table}[htbp]
    \caption{Temperature \& Humidity Sample Data}
    \label{tab:temp_humidity}
\begin{tabular}{@{}lrr@{}}
\toprule
\textbf{Time} & \textbf{Temperature ($^\circ$C)} & \textbf{Humidity (\%)} \\
\midrule
2025-10-09 00:00 & 24.9 & 71.1 \\
2025-10-09 00:05 & 24.7 & 70.9 \\
2025-10-09 00:10 & 24.7 & 71.3 \\
2025-10-09 00:15 & 24.8 & 71.3 \\
2025-10-09 00:20 & 24.4 & 71.5 \\
2025-10-09 00:25 & 24.4 & 72.2 \\
\bottomrule
\end{tabular}
\end{table}

\paragraph{Domain Specific Files}
After setting the environment variable \texttt{DATA\_TYPE = "temp\_humidity"} in the \textit{config.py} file, we provide the following files, which are stored inside the \textit{temp\_humidity folder}, to the LLM:

\begin{itemize}
\item \textit{sample\_data.csv}: A subset of temperature and humidity data (as shown in table \ref{tab:temp_humidity}).
    \item \textit{meta\_data.json}: A JSON file defining units (e.g., Celsius, Percentage) and column descriptions.
    \item \textit{context.txt}: A text file describing data collection goals, such as monitoring thermal comfort levels.
    \item \textit{tasks\_list.json} (optional). The complete metadata schema, content, and tasks\_list are available in the \texttt{/data/temp\_humidity} directory of the GitHub repository~\cite{github}.
\end{itemize}

\paragraph{Pipeline Steps}
Along with the domain-specific files, the resource statistics of the edge device are sent to the LLM. For this use case, the LLM returns tasks such as ``overheating\_degree\_minutes\_temperature'' and ``wet\_bulb\_temperature\_stull''. After receiving the list of tasks in \textit{new\_task.json}, the \textit{task\_code\_generator.py} generates the respective code, such as \textit{overheating\_degree\_minutes\_temperature.py}. Once generated, the code is validated by the \textbf{\textit{validator.py}}, producing the \textit{validator\_summary.json} file. If no error arises, the code is stored in the intelligence repository. Finally, the Scheduler executes the validated code, generates the log file, and sends the result for publication.
\vspace{0.5em}
\subsubsection{Use Case 3: Soil Monitoring}
  
\paragraph{Objective \& Data Source}
This use case targets the agricultural area by monitoring soil conditions at various depths. The objective is to analyze moisture retention and temperature to optimize irrigation. The dataset contains daily records of Soil Temperature ($^\circ$C) and Soil Moisture ($m^3/m^3$) measured at multiple depths (0cm, 6cm, 18cm, 54cm, etc.). Table {tab:soil\_data} presents a subset of the raw data.

\begin{table}[htbp]
\caption{Soil Temperature and Moisture Sample Data}\label{tab:soil_data}
\begin{tabular}{@{}lrrrrr@{}}
\toprule
\textbf{Date} & \textbf{Temp (0cm)} & \textbf{Temp (18cm)} & \textbf{Temp (54cm)} & \textbf{Moisture (0--1cm)} & \textbf{Moisture (9--27cm)} \\
\midrule
2025-12-10 18:30:00+00:00 & 8.054     & 19.804 & 21.604 & 0.106 & 0.150 \\
2025-12-10 19:30:00+00:00 & 7.554     & 19.504 & 21.604 & 0.107 & 0.151 \\
2025-12-10 20:30:00+00:00 & 7.4040003 & 19.204 & 21.604 & 0.107 & 0.151 \\
2025-12-10 21:30:00+00:00 & 6.854     & 18.854 & 21.604 & 0.108 & 0.151 \\
2025-12-10 22:30:00+00:00 & 6.304     & 18.554 & 21.554 & 0.108 & 0.151 \\
2025-12-10 23:30:00+00:00 & 6.354     & 18.254 & 21.554 & 0.107 & 0.153 \\
\bottomrule
\end{tabular}
\end{table}

\paragraph{Domain Specific Files}
After setting the environment variable \texttt{DATA\_TYPE = "soil"} in the \textit{config.py} file, the system utilizes the following files located in the \textit{soil} folder:
\begin{itemize}
    \item \textit{sample\_data.csv}: It is shown in the table \ref{tab:soil_data}.
    \item \textit{meta\_data.json}: Defines the units for temperature ($^\circ$C) and moisture content.
    \item context\.txt: This file describes the analytical goals, such as irrigation efficiency or root zone health assessment risks. 
    \item \textit{tasks\_list.json} (optional).
\end{itemize}


\paragraph{Pipeline Steps}
The resource statistics and domain files are sent to the LLM, which generates specific agricultural tasks. For examples, the LLM proposes tasks such as ``root\_zone\_health'' and ``temperature\_moisture\_trends''. The \textit{task\_code\_generator.py} creates the corresponding Python scripts (e.g., \textit{root\_zone\_health.py} and \textit{temperature\_moisture\_trends.py}), which are then validated. The final validated code is executed by the Scheduler to generate insight logs and visualization plots for the stakeholders.

\vspace{0.5em}
\subsubsection{Use Case 4: Wind Monitoring}

\paragraph{Objective \& Data Source}
This use case focuses on renewable energy from wind by analyzing its speed and direction at different altitudes. The objective is to evaluate wind speed gradients and directions. The dataset comprises measurements of Wind Speed (m/s) and Wind Direction ($^\circ$) taken at heights of 10m, 80m, 120m, and 180m, along with surface wind gusts at 10m. Table \ref{tab:wind_data} presents a subset of the raw data.

\begin{table}[htbp]
\caption{Wind Speed \& Direction Sample Data}\label{tab:wind_data}
\begin{tabular}{@{}lrrrrr@{}}
\toprule
\textbf{Date} & \textbf{Speed (10m)} & \textbf{Speed (120m)} & \textbf{Dir (10m)} & \textbf{Dir (120m)} & \textbf{Gusts (10m)} \\
\midrule
2025-12-10 18:30:00+00:00 & 2.968636  & 5.692099  & 284.03625 & 34.69522  & 3.96 \\
2025-12-10 19:30:00+00:00 & 3.319036  & 6.9527545 & 282.5288  & 21.25058  & 4.68 \\
2025-12-10 20:30:00+00:00 & 3.671294  & 7.172949  & 281.3099  & 17.52567  & 4.68 \\
2025-12-10 21:30:00+00:00 & 3.2599385 & 7.172949  & 276.3401  & 17.52567  & 4.68 \\
2025-12-10 22:30:00+00:00 & 3.2599385 & 6.9527545 & 276.3401  & 21.25058  & 4.68 \\
2025-12-10 23:30:00+00:00 & 3.4152596 & 8.766573  & 288.43503 & 19.179106 & 4.68 \\
\bottomrule
\end{tabular}
\end{table}

\paragraph{Domain Specific Files}
After setting the environment variable \texttt{DATA\_TYPE = "wind\_energy"} in the \textit{config\.py} file, the system utilizes the following files located in the \textit{wind} folder:

\begin{itemize}
    \item \textit{sample\_data.csv}: It is shown in the table \ref{tab:wind_data}.
    \item \textit{meta\_data.json}: Defines the units for speed (m/s) and direction in degrees. It also explicitly maps the column names to their respective measurement heights.    
    \item \textit{context.txt}: Describes the analytical goals, such as calculating wind shear exponents, generating wind rose diagrams, and analyzing gust factors.
\end{itemize}

Detailed information about the sample data, meta\_data, and context about the wind are available in the directory \texttt{/data/wind} of the same GitHub repository\cite{github}.

\paragraph{Pipeline Steps}
The resource statistics and domain-specific files are sent to the LLM for task generation. For this dataset, the LLM proposes tasks such as ``altitude\_shear\_profile\_analyzer'' and ``wind\_rose\_generator''. The \textit{task\_code\_generator.py} creates the corresponding Python scripts, which are then validated to ensure they can handle the multi-column altitude data correctly. The final validated code is executed by the Scheduler to generate insight logs and visualization plots.

\subsection{Performance evaluation metrics}
To evaluate the effectiveness of our proposed system across different models on different use cases on resource-constrained devices, we analyze using the following key metrics:

\begin{enumerate}
    \item Latency: It measures the time taken by the system and reports in either milliseconds (ms) or seconds (sec).
    \begin{itemize}
        \item \textbf{Task Generation Latency} ($L_{s1}$): Time required to generate a task list from the input environmental data (Step 1).
        \item \textbf{Code Generation} ($L_{s2}$): Time taken by the LLM to generate executable code (Step 2).
        \item \textbf{Validation Latency} ($L_{s3}$): Time taken by the system by the validator to verify correctness (Step 3).
        \item \textbf{Execution Latency} ($L_{s4}$): Time taken by the scheduler to execute validated code and generate JSON outputs (Step 4).
        \item \textbf{End-to-End Latency} (L): Total time from user prompt to final JSON output as mentioned in Equation \ref{eq:total_latency}:
    \end{itemize}
    
    \begin{equation} \label{eq:total_latency}
    L = L_{\text{s1}} + L_{\text{s2}} + L_{\text{s3}} + L_{\text{s4}}
    \end{equation}

    \item Resource Utilization: It refers to how edge devices consume their resources, such as CPU, Memory, and Network I/O, during the execution of the system. In an edge environment, this metric plays a crucial role in real-world deployment, as it helps to verify that the pipeline is deployable on resource-constrained devices. Higher resource usage leads to overheating or system failure, while lower resource usage keeps the edge device temperature at a minimal rise, which helps in the long run without risking hardware failure. We have collected resource metrics every 5 seconds throughout the pipeline execution for each use case and model, multiple times. To calculate the CPU and Memory in percentage, we have use the equation \ref{eq:resource} and the results are shown in Figures \ref{fig:avg_cpu_res} and \ref{fig:avg_mem_res}.

    \begin{equation}
        \text{AvgRun}_{m,s,r,p} = \frac{1}{N_r}\sum_{i=1}^{N_r} x_{m,s,r,i,p}
    \end{equation}
    \begin{equation}\label{eq:resource}
        \text{RU}_{m,s,r,p} = \frac{1}{R}\sum_{i=1}^{R} \text{AvgRun}_{m,s,r,p}
    \end{equation}

    where:
    \begin{itemize}
        \item $x_{m,s,r,i,p}$ denotes the $i$-th individual sample recorded for a particular model $m$, at step $s$, during run $r$.
        \item $N_r$ denotes the total number of samples recorded for a unique run.
        \item $\text{AvgRun}_{m,s,r,p}$ demoted intermediate average calculated for a specific run for a each model $m$ at each step $s$.
        \item $R$ denotes total number of runs performed for a specific model $m$.
        \item $\text{RU}_{m,s,r,p}$ denotes the final average of total runs $R$ for model $m$ at step $s$.
        \item $p \in \{\text{CPU, Memory}\}$
    \end{itemize}
    
    \item Reliability: It is defined as the system's ability to generate correct and executable code. In the proposed framework, the validator component in Step 3 ensures that no false code reaches the scheduler in Step 4 (as shown in Figure \ref{fig:sbs}). A code is considered validated if it passes through the validator either on the initial run or after one or two correction attempts. Let $P_0$ represent the set of codes that passed validation on the initial attempt, $P_{12}$ represent the set of codes that passed after one or two attempts, and $F$ represents the code that failed after two attempts. The system reliability ($R_{sys}$) is calculated as the ratio of all the validated code to the total number of codes received for validation, given in Equation \ref{eq:reliable}:

    \begin{equation} \label{eq:reliable}
        R_{sys} = \frac{P_0 + P_{12}}{P_0 + P_{12} + F}
    \end{equation} 
    The reliability of the proposed system is shown in Figure \ref{fig:val_result}, with a detailed discussion in the later sections.

    \item Throughput: It is defined as the number of tokens processed or generated per second. We measure throughput in terms of token processing and token generation, given by Equations \ref{eq:ptps} and \ref{eq:ctps}, respectively.
    
    \begin{equation}\label{eq:ptps}
        \text{prompt\_token\_per\_sec} = \frac{prompt\_token}{llm\_duration\_sec}
    \end{equation}
    \begin{equation}\label{eq:ctps}
        \text{completion\_token\_per\_sec} = \frac{completion\_token}{llm\_duration\_sec}
    \end{equation}
    
\end{enumerate}

\begin{table}[htbp]
\caption{LLM (Using OLLAMA) Inference Metrics Definitions}\label{tab:llm_metrics_def}
\begin{tabular}{@{}lp{11cm}@{}}
\toprule
            \textbf{Metric Name} & \textbf{Description} \\ 
\midrule        
            \texttt{response\_token/s} & The rate at which the model generates output tokens (Generation Speed). \\ 
            \texttt{prompt\_token/s} & The rate at which the model processes the input prompt (Ingestion Speed). \\ 
            
            \texttt{total\_duration} & The absolute total time elapsed for the entire request (Load + Prompt + Eval). \\ 
            \texttt{load\_duration} & Time taken to load the model weights into memory/VRAM before processing starts. \\ 
            \texttt{prompt\_eval\_duration} & Time spent specifically calculating embeddings/attention for the user's input prompt. \\ 
            \texttt{eval\_duration} & Time spent solely on generating the new response tokens. \\ 
            
            \texttt{prompt\_tokens} & The number of tokens in the user's input (Question + Context). \\ 
            \texttt{prompt\_eval\_count} & The number of input tokens that were actually processed/evaluated during this run. \\ 
            \texttt{completion\_tokens} & The number of new tokens generated by the model as the answer. \\ 
            \texttt{eval\_count} & Same as completion\_tokens; the count of tokens evaluated during generation. \\ 
            \texttt{total\_tokens} & The sum of prompt\_tokens and completion\_tokens (Total Context used). \\ 
            
            \texttt{approximate\_total} & A human-readable formatting of the total duration (e.g., "16s"). \\ 
\botrule
\end{tabular}

\end{table}

To maintain uniform comparison, we have included models with fewer than 10 billion parameters, allowing the edge device to run smoothly.

\subsection{Numerical Results}\label{sec:res}

This section presents an evaluation of the proposed work using four different environmental datasets and eight models. All models are run 10 times per dataset on Raspberry Pis. Since each step is interconnected, starting from generating a list of tasks in JSON to executing scripts, suppose a model fails at any step due to LLM timeout or invalid file formats received or fails to generate correct code, then for that particular run, the model results will be empty or zero or removed from the result graph. In the Figure \ref{fig:avg_cpu_res} and \ref{fig:avg_mem_res}, M1 corresponds to codegemma:7b, M2 to codellama:7b, M3 to deepseek-coder:6.7b, M4 to gemma3:4b, M5 to llama3.1:latest, M6 to mistral:7b, M7 to phi3:3.8b, and M8 to qwen2.5-coder:7b.

\subsubsection{Average CPU \& Memory usage across four cases}
\begin{figure}[htbp]
    \centering
    \includegraphics[width=1\linewidth]{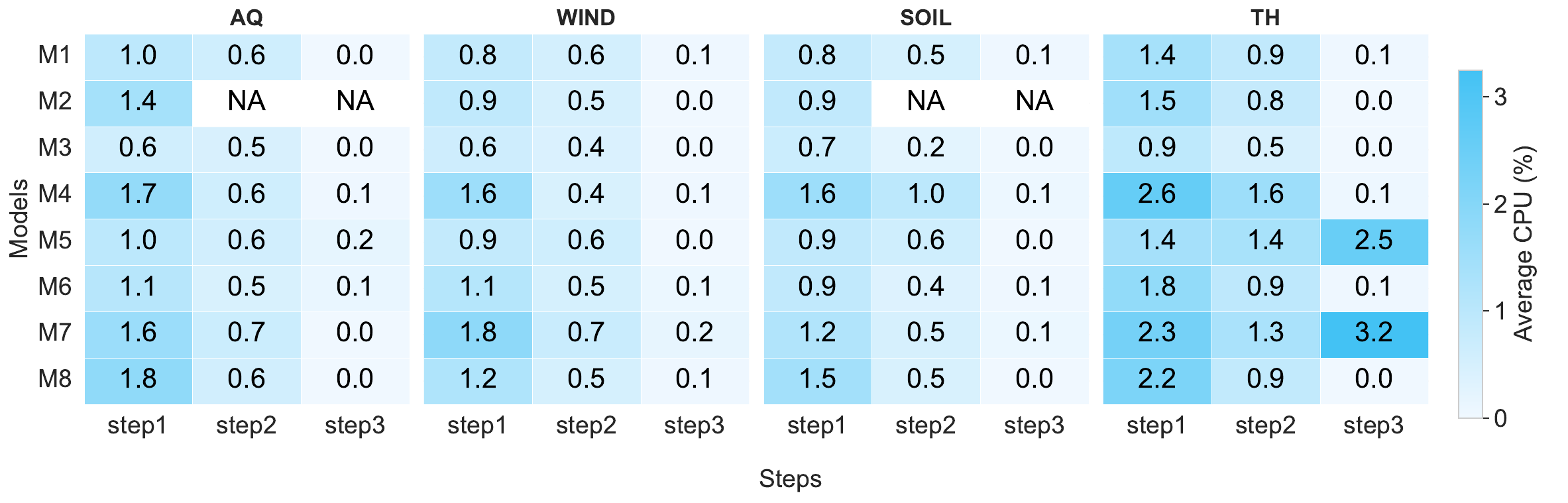}
    \caption{Average CPU usage of edge devices.}
    \label{fig:avg_cpu_res}
\end{figure}

Figure \ref{fig:avg_cpu_res} shows the average CPU utilization in percentage (\%) per model for the first three steps of the pipeline as shown in the \ref{fig:sbs}. Across all use cases \& steps the overall CPU usage remains below 3\% except M7 (phi3:3.8b) model which show 3.2 \%. Step 1 (Task Generation) exhibits the highest CPU load, ranging from 0.6 to 2.6 \%. In Step 2 (Code Generation), CPU usage drops across all models to 0.2-1.6 \%. Step 3 (Validation) shows less than ($\approx$0.001\%) amount of CPU usage with an exception in TH dataset for models M5 \& M7. We noted that model M2 for AQ \& SOIL dataset doesn't record any measurement and model M3 takes the least CPU \% across all use cases at each step.

\begin{figure}[htbp]
    \centering
    \includegraphics[width=1\linewidth]{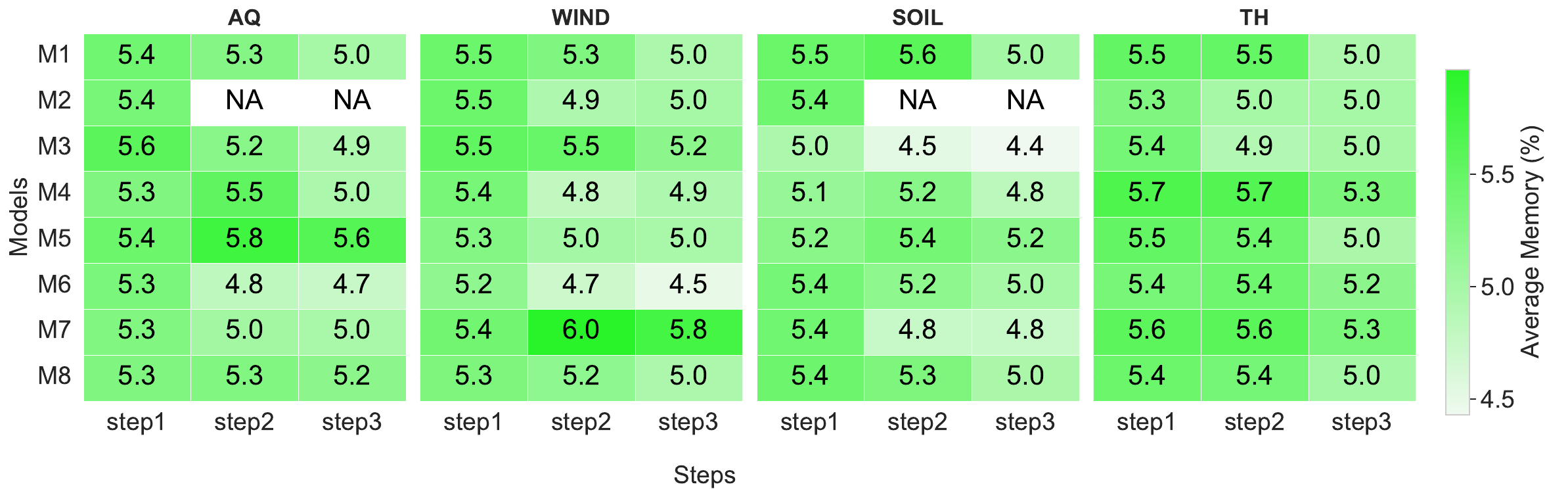}
    \caption{Average memory usage of edge devices.}
    \label{fig:avg_mem_res}
\end{figure}

Unlike CPU usage, Memory usage remains stable during each step within the range of 4.5 - 6.0 \%, as shown in the Figure \ref{fig:avg_mem_res}. The model M7 (phi3\_3.8b), exhibits the highest 6.0 \% memory utilization in the Wind dataset at step 2, while model M3 (deepseek-coder:6.7b) recorded the lowest memory usage at step 1 (5.0\%), step 2 (4.5 \%) and step 3 (4.4 \%) for SOIL dataset. Grey cells in Figures \ref{fig:avg_cpu_res} and \ref{fig:avg_mem_res} indicate that the corresponding models failed to reach pipeline step 2 (Fig. \ref{fig:sbs}) for the given dataset.

\subsubsection{ Throughput across four cases}
\begin{figure}[htbp]
    \centering
    \includegraphics[width=\linewidth, height=6cm, keepaspectratio]{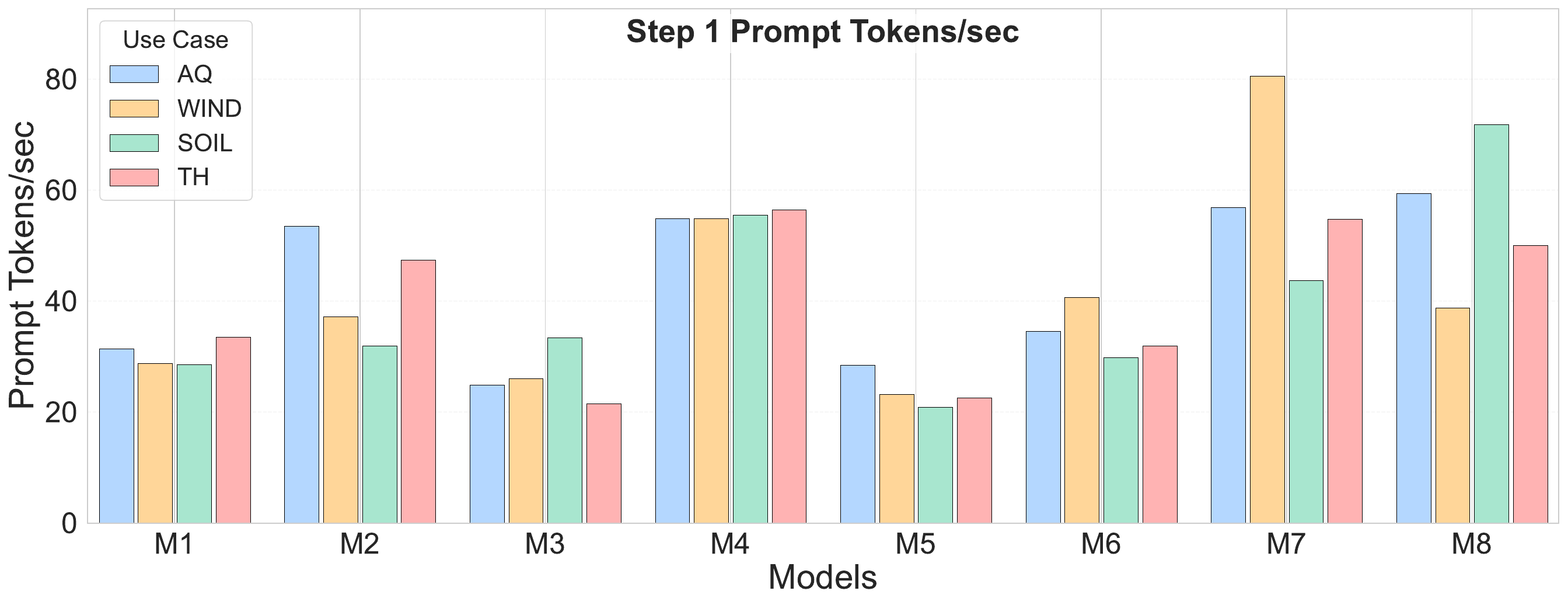}
        \caption{Prompt token per sec. Number of prompt tokens per sec in step 1}
        \label{fig:ptps_step1}
\end{figure}

\begin{figure}[htbp]
    \centering
    \includegraphics[width=\linewidth, height=6cm, keepaspectratio]{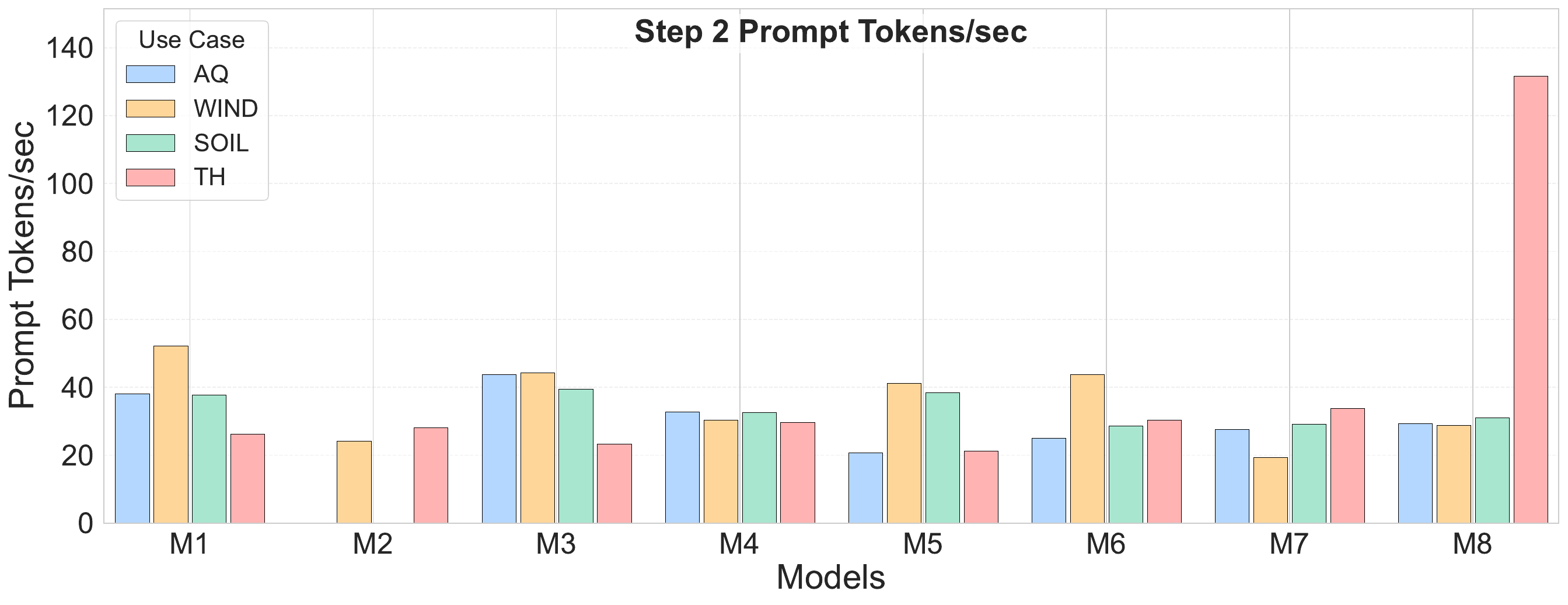}
        \caption{Prompt token per sec. Number of prompt tokens per sec in step 2} 
        \label{fig:ptps_step2}
\end{figure}

\begin{figure}[htbp]
    \centering
    \includegraphics[width=\linewidth, height=6cm, keepaspectratio]{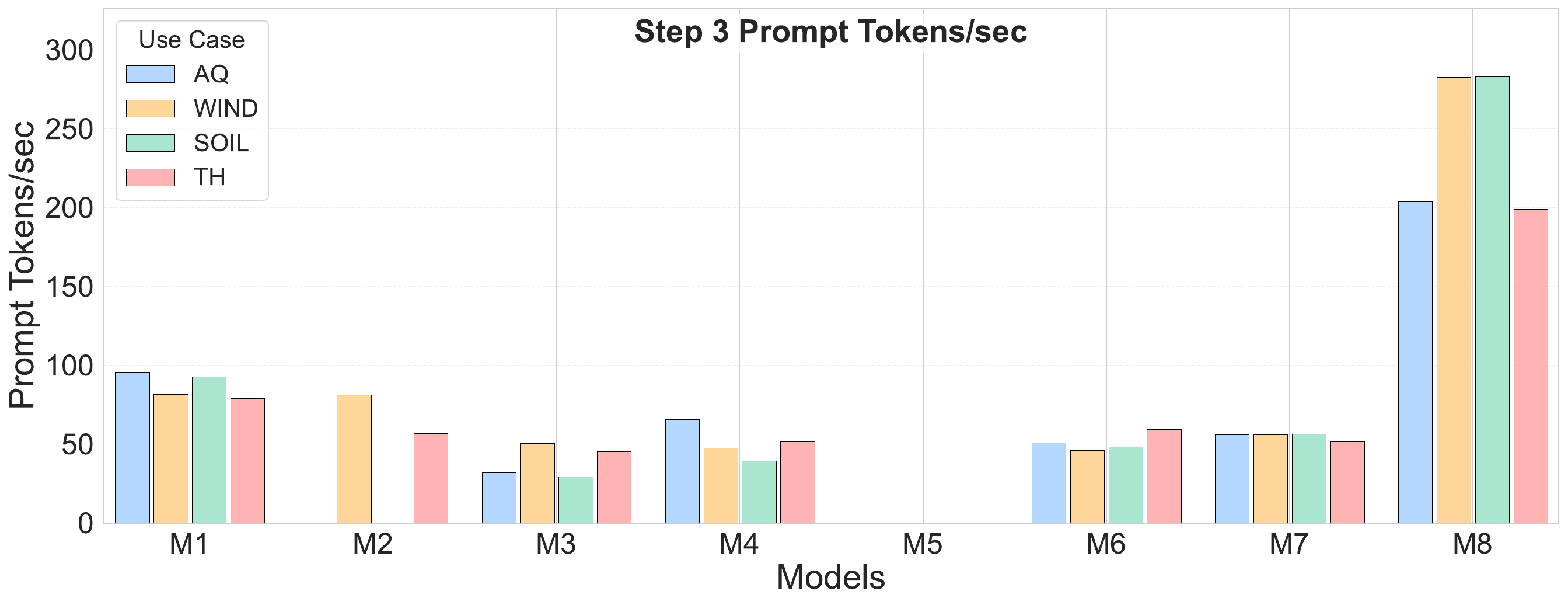}
        \caption{Number of prompt tokens per sec in step 3. See Table \ref{tab:ollama_models} for model ID definitions.}
        \label{fig:ptps_step3}
\end{figure}

Fig.~\ref{fig:ptps_step1},~\ref{fig:ptps_step2},~\ref{fig:ptps_step3} illustrate the variation in prompt token throughput across models, steps, and use cases. The results show that qwen2.5\_coder\_7b outperforms all other models, achieving the highest overall throughput across the three steps, with total prompt token rates reaching $\approx$ 300-400 tokens/s, depending on the use case. Step 2 throughput generally ranges from $\approx$ 20-80 tokens/s across all models. In this step, we note that gemma3\_4b and phi3\_3.8b achieve the highest performance ($\approx$ 50-80 tokens/s), whereas llama3.1\_latest exhibits the lowest Step 1 throughput, remaining below 30 tokens/s for all use cases. For Step 2, throughput typically ranges between 40 and 90 tokens/s. deepseek\_coder\_6.7b and codegemma\_7b demonstrate comparatively strong Step 2 performance, while codellama\_7b shows no measurable throughput for the evaluated AQ and SOIL use cases. For most models, Step 3 contributes the largest prompt tokens per second; for example, codegemma\_7b and qwen2.5\_coder\_7b which achieves $\approx$ 200 tokens/s at this step. In contrast, llama3.1\_latest does not produce results for Step 3 across all the evaluated scenarios.

\begin{figure}[htbp]
    \centering
    \includegraphics[width=\linewidth, height=6cm, keepaspectratio]{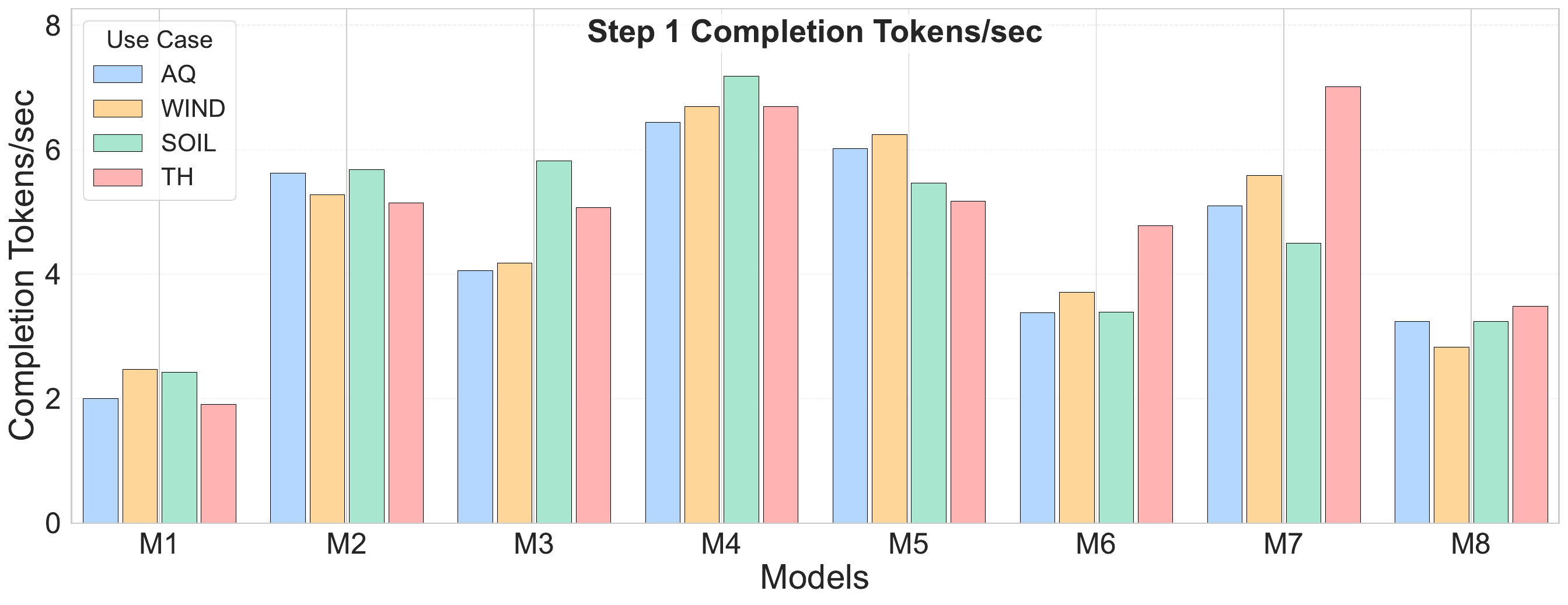}
        \caption{Completion token per sec. Number of completion tokens per sec in step 1}
        \label{fig:ctps_step1}

\end{figure}

\begin{figure}[htbp]
    \centering
    \includegraphics[width=\linewidth, height=6cm, keepaspectratio]{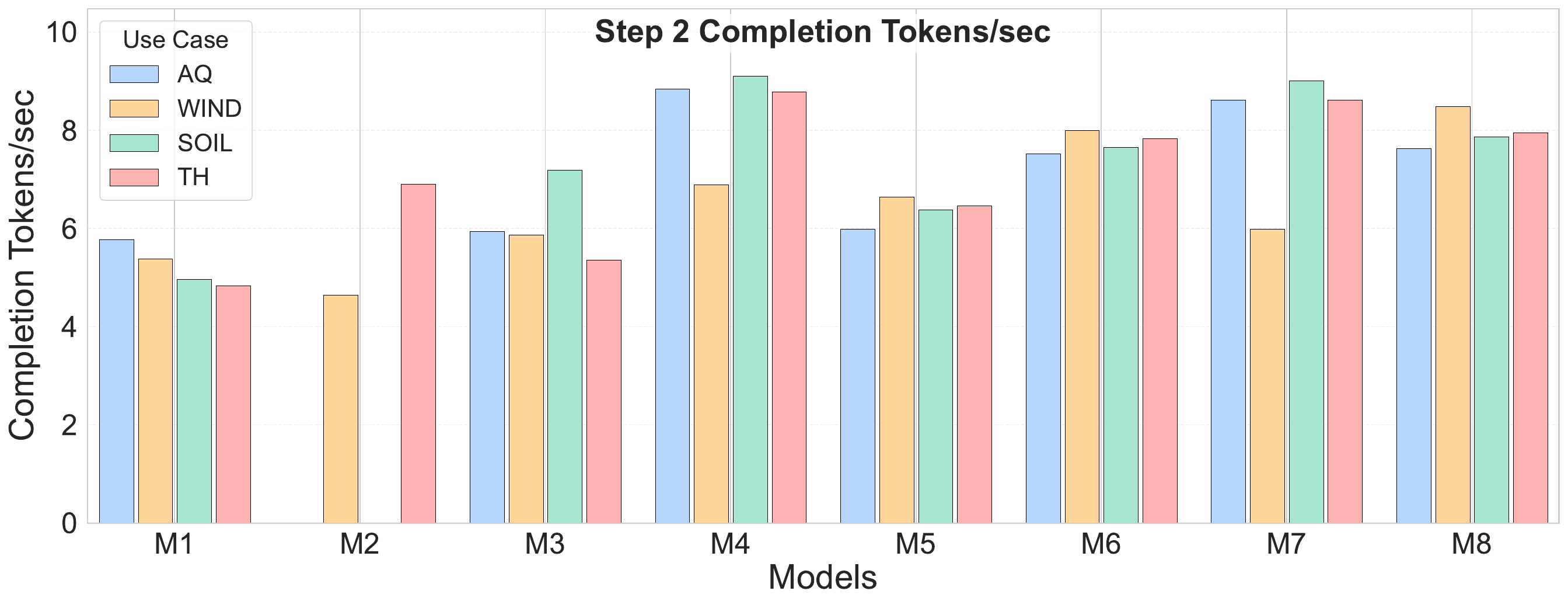}
        \caption{Completion token per sec. Number of completion tokens per sec in step 2} 
        \label{fig:ctps_step2}
\end{figure}

\begin{figure}[htbp]
    \centering
    \includegraphics[width=\linewidth, height=6cm, keepaspectratio]{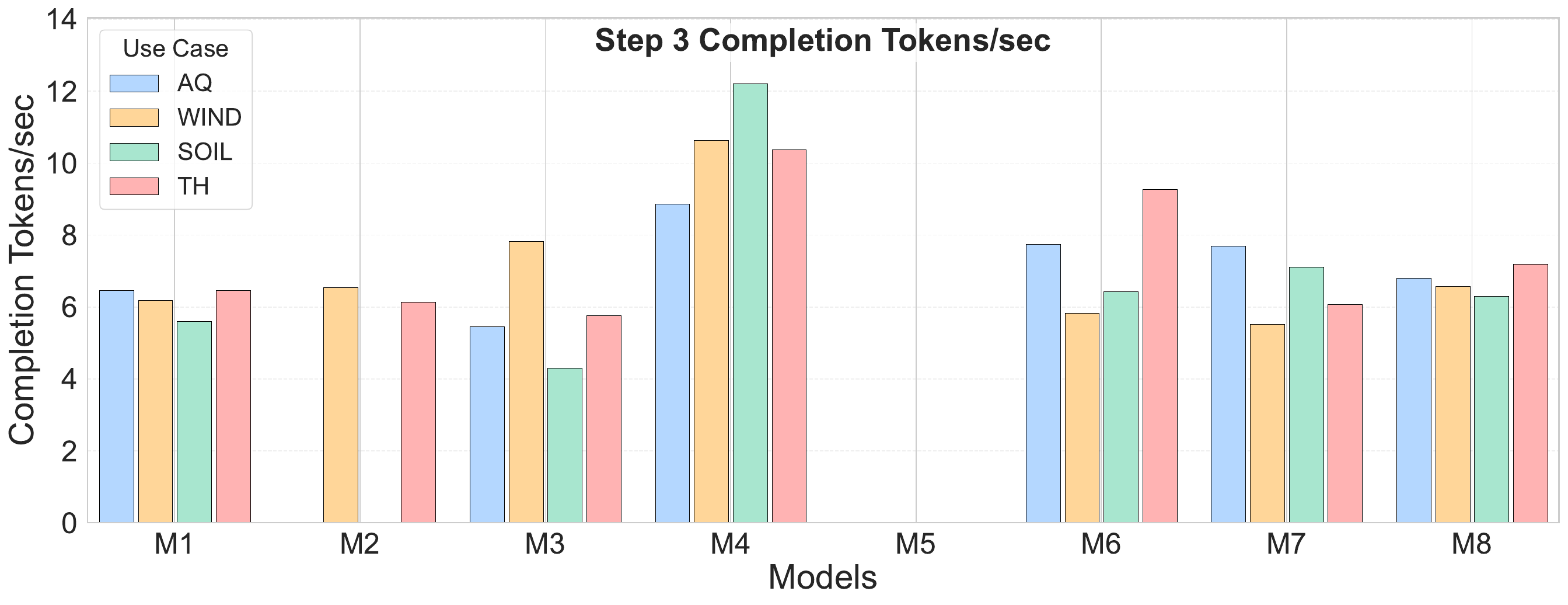}
        \caption{Completion token per sec. Number of completion tokens per sec in step 3. See Table \ref{tab:ollama_models} for model ID definitions.}
        \label{fig:ctps_step3}
\end{figure}

In contrast to prompt throughput, completion rates are lower overall, remaining below 30 tokens/s for all models, as shown in Fig.~\ref{fig:ctps_step1},~\ref{fig:ctps_step2},~\ref{fig:ctps_step3}, which presents the variation in completion token throughput across models, steps, and use cases. Among the evaluated models, we observe that gemma3\_4b completes inference more quickly by achieving the highest aggregate completion throughput, reaching $\approx$ 25-30 tokens/s. For Step 1, completion throughput typically ranges from $\approx$ 3 to 8 tokens/s, with the highest values observed for gemma3\_4b, phi\_3.8b, llama3.1\_latest, and codellama\_7b ($\approx$ 6-7 tokens/s), whereas codegemma\_7b and qwen2.5\_coder\_7b achieve the lower values of this range. Step 2 shows a moderate increase in throughput, with values generally between 5 and 10 tokens/s. In this step, gemma3\_4b, mistral\_7b, and qwen2.5\_coder\_7b demonstrate comparatively strong performance, exceeding 8 tokens/s, while codellama\_7b exhibits no measurable performance in the AQ and SOIL use cases. For Step 3, gemma3\_4b achieves the highest throughput ($\approx$ 11 tokens/s), followed by mistral\_7b and qwen2.5\_coder\_7b, each reaching $\approx$ 8 tokens/s. In contrast, llama3.1\_latest does not produce measurable results for Step 3 across all use cases. 

If any model fails to generate a valid output at any step, subsequent steps are not executed. For example, codellama:7b fails to produce a list of tasks in the required format (i.e., JSON structure) or to return an empty response in Step 1, which prevents the execution of Step 2. Similarly, llama3.1\_latest shows no throughput in step 3 because the pipeline terminated at step 2 due to LLMs returning invalid JSON or an empty response. Is summary, we observe that token consumption for prompt is higher than completion, as the prompt requires sample data, metadata, context, resource statistics, and the previous task list, all of which are provided to the LLM. On the other hand, completion consists of the generated task list or Python code in JSON format.

\subsubsection{End-to-End Latency across four cases}
\begin{figure}[htbp]
    \centering
    \includegraphics[width=0.95\linewidth, height=10cm, keepaspectratio]{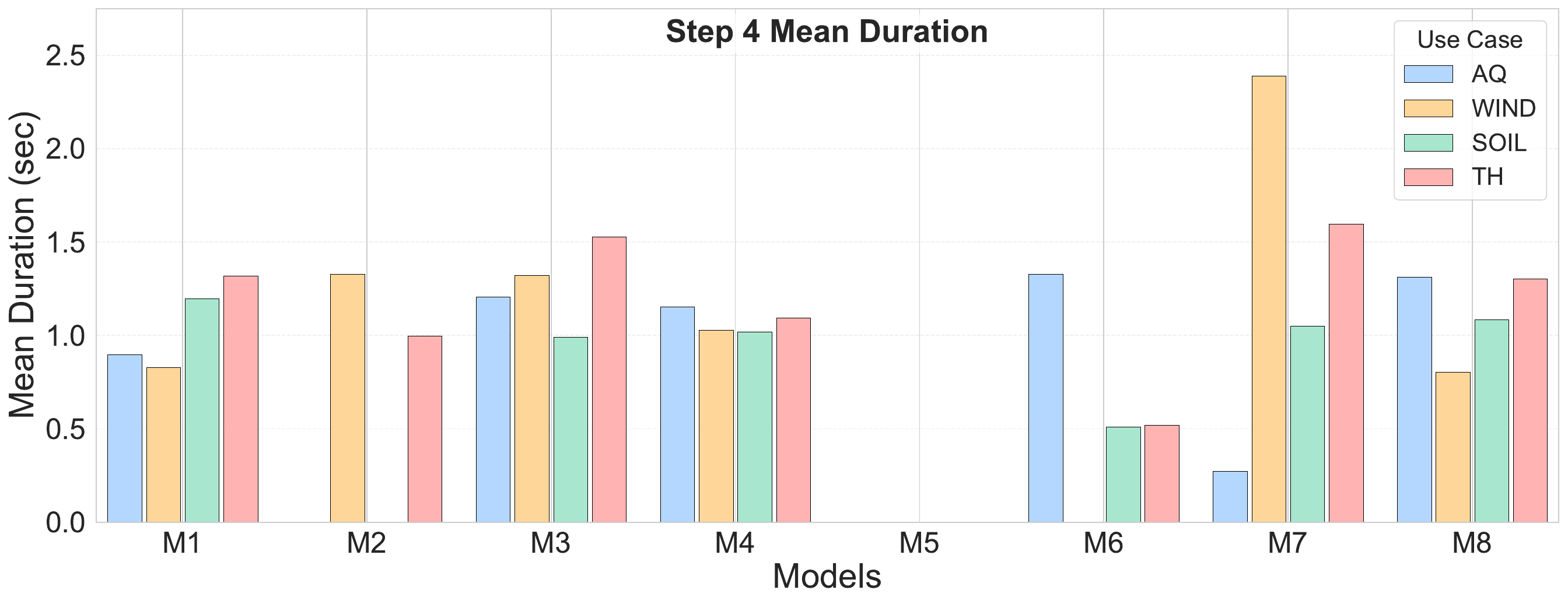}
    \caption{End-to-End Latency for four use cases (See Table \ref{tab:ollama_models} for model ID definitions)}
    \label{fig:latency}
\end{figure}

Fig.~\ref{fig:latency} presents the end-to-end latency across models, decomposed by individual steps and evaluated over four use cases (AQ, WIND, SOIL, and TH). The results show variation in total latency across models, driven by Steps 2 and 3. Overall, gemma3\_4b achieves faster overall inference and exhibits the lowest end-to-end latency among the evaluated models, primarily due to lower Step 3 execution times, with total durations generally remaining below 300–350 s across use cases, except in Step 2 of WIND use case. In contrast, mistral\_7b and deepseek\_coder\_6.7b show the highest latency, reaching $\approx$ 800–1000 s, as example in SOIL use case.

Step 1 latency remains low across all models, generally below 150 s, with the exception of deepseek\_coder\_6.7b, which exhibits higher Step 1 durations. We note that Step 2 introduces a notable increase in latency, particularly for deepseek\_coder\_6.7b and gemma3\_4b, where it exceeds 300 s. Step 3 is the main contributor to total latency for mistral\_7b, with more than 50\% of the overall execution time, with extreme cases exceeding 500 s, specifically for AQ, WIND, and SOIL. In contrast, llama3.1\_latest does not report measurable results for Step 2 in the AQ and SOIL use cases, and codellama\_7b does not report measurable results for Step 3 across all evaluated use cases. Step 4 shows the lowest latency across all models, generally ranging from 0.1 to 1.6 s, except for phi3\_3.8b on the wind dataset, which takes 2.3 s. In the Fig. \ref{fig:latency}, any models showing no latency at a particular step show that they fail to complete the previous step due to returning invalid results.

\begin{figure}[htbp]
    \centering
    \includegraphics[width=\linewidth, height=6cm, keepaspectratio]{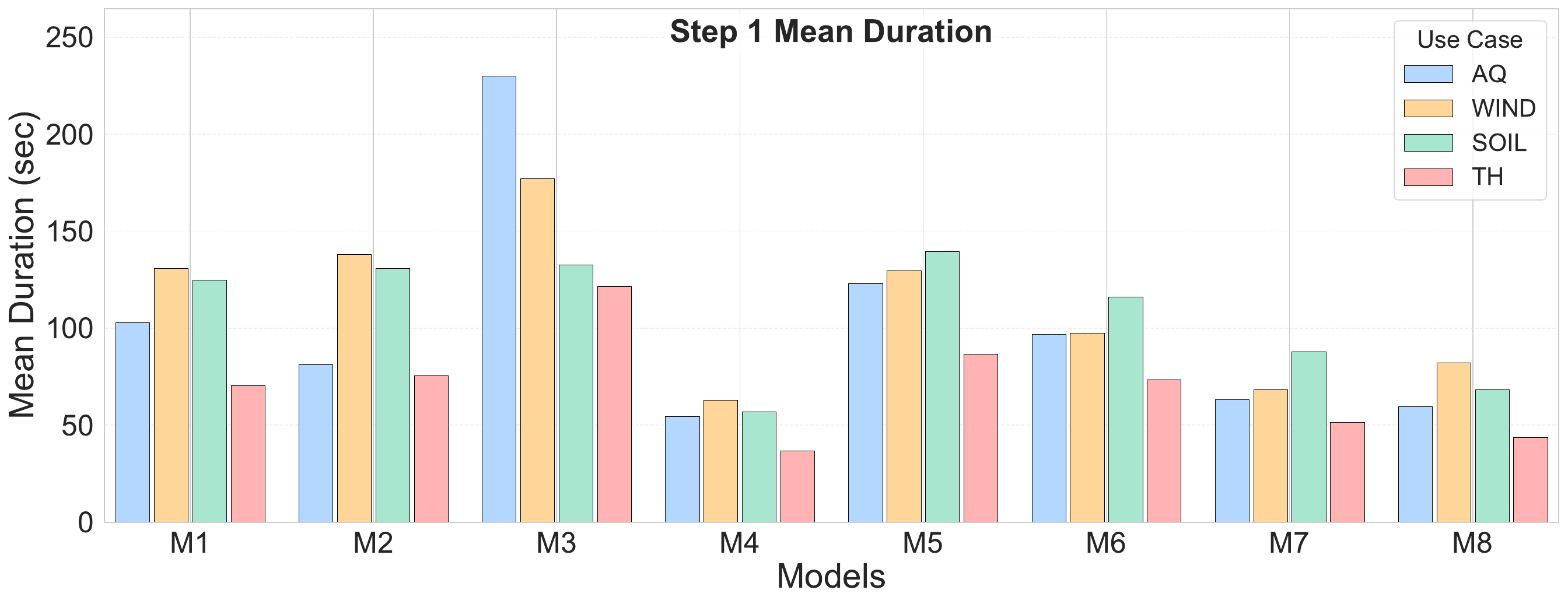}
        \caption{End-to-End Latency. Mean Duration in step 1}
        \label{fig:late_step1}
\end{figure}
    
\begin{figure}[htbp]
    \centering
    \includegraphics[width=\linewidth, height=6cm, keepaspectratio]{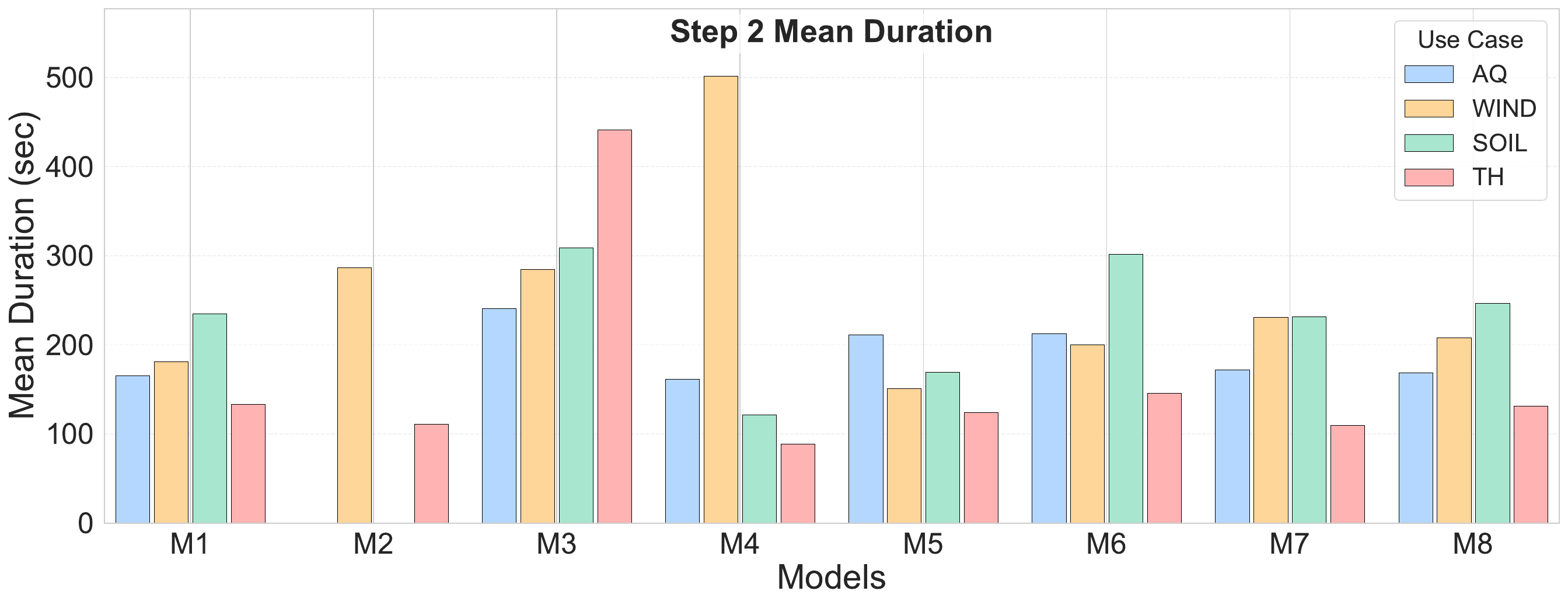}
        \caption{End-to-End Latency - Mean Duration in step 2} 
        \label{fig:late_step2}
\end{figure}
    
\begin{figure}[htbp]
    \centering
    \includegraphics[width=\linewidth, height=6cm, keepaspectratio]{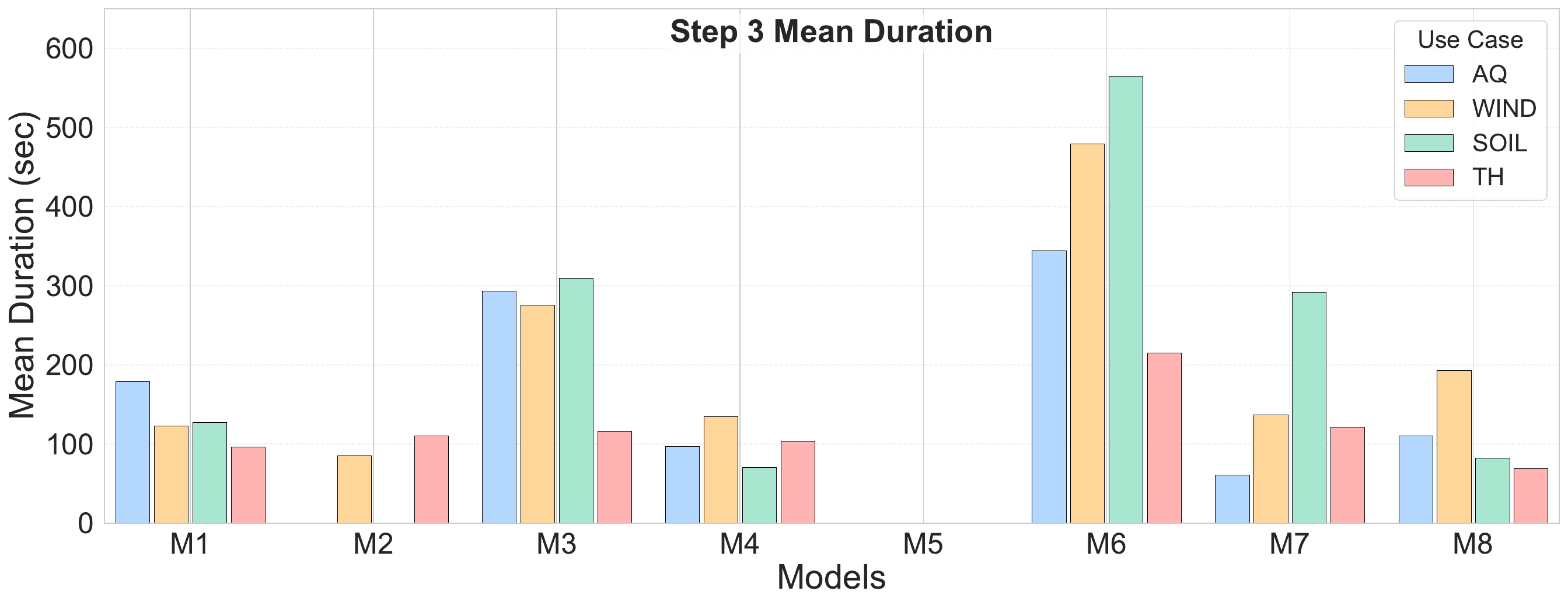}
        \caption{End-to-End Latency - Mean Duration in step 3. See Table \ref{tab:ollama_models} for model ID definitions}
        \label{fig:late_step3}

\end{figure}

\subsubsection{Complete Pipeline}
\begin{figure}[htbp]
    \centering
    \includegraphics[width=0.95\linewidth]{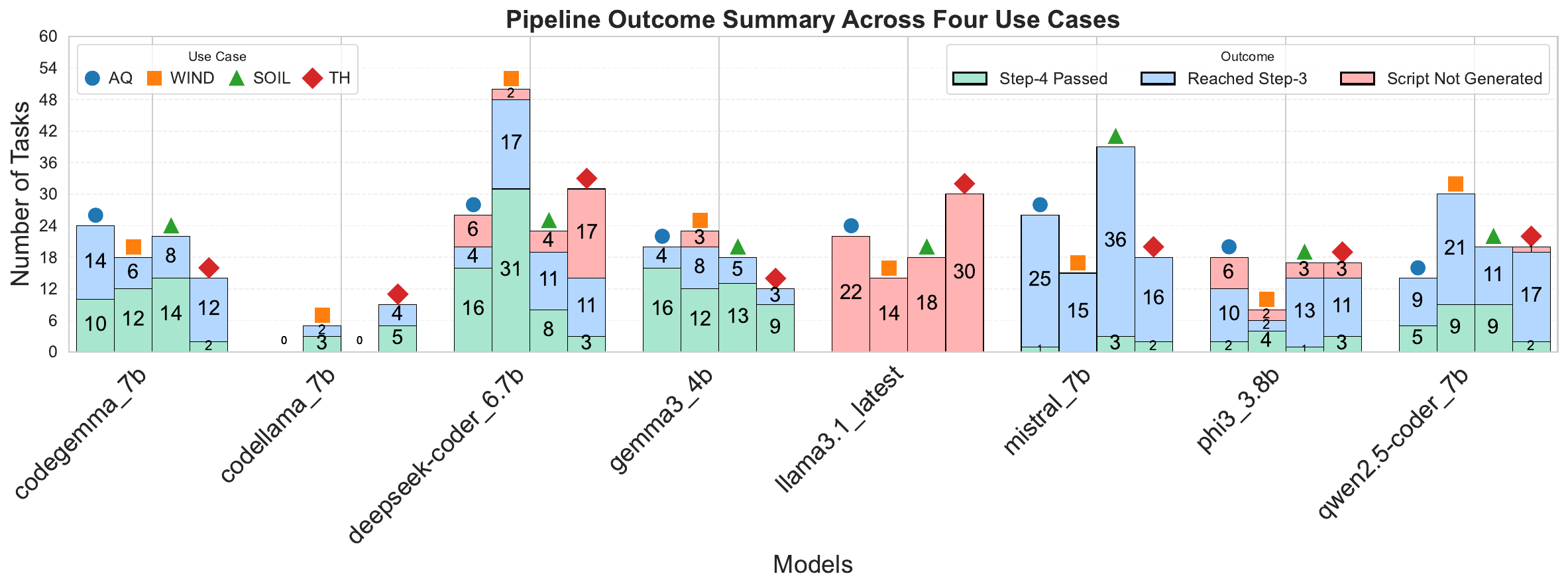}
    \caption{Complete Pipeline for four cases}
    \label{fig:pipeline_result}
\end{figure}

Figure \ref{fig:pipeline_result} presents the end-to-end pipeline outcome across four environment datasets for eight models. Overall, the results show a clear relation between model type and pipeline completion success rate. The code-based models, such as deepseek\_coder\_6.7b, codegemma\_7b, and qwen2.5\_coder\_7b, outperform the general-purpose models like mistral\_7b and phi3\_3.8b in terms of completion success rate in reaching Step 4 (execution). Among all the models, deepseek-coder-6.7b achieves the highest number of successfully executed scripts across most datasets. Notably, the only general-purpose model gemma3\_4b, despite having fewer parameters, performs strongly and ranks second in several use cases. For all use cases, an anomaly with the mistral\_7b model is that it keeps generating incorrect scripts, gets stuck at step 3, and only a few scripts reach Step 4, as shown in \ref{fig:sbs}, while llama3.1\_latest fails to generate valid scripts. For a clear understanding of the figure, consider the example where the deepseek-coder-6.7b model on the WIND dataset generates 50 total tasks across 10 runs in step 1. Out of these, only 2 tasks fail to generate valid scripts in step 2, and the remaining 48 scripts proceed to step 3 for validation. Of these 48, only 17 fail as marked by the validator in step 3 and are discarded. Finally, the remaining 31 scripts are ready for execution by the scheduler in step 4.

\subsubsection{Validation Success Rate}
\begin{figure}[htbp]
    \centering
    \includegraphics[width=0.95\linewidth]{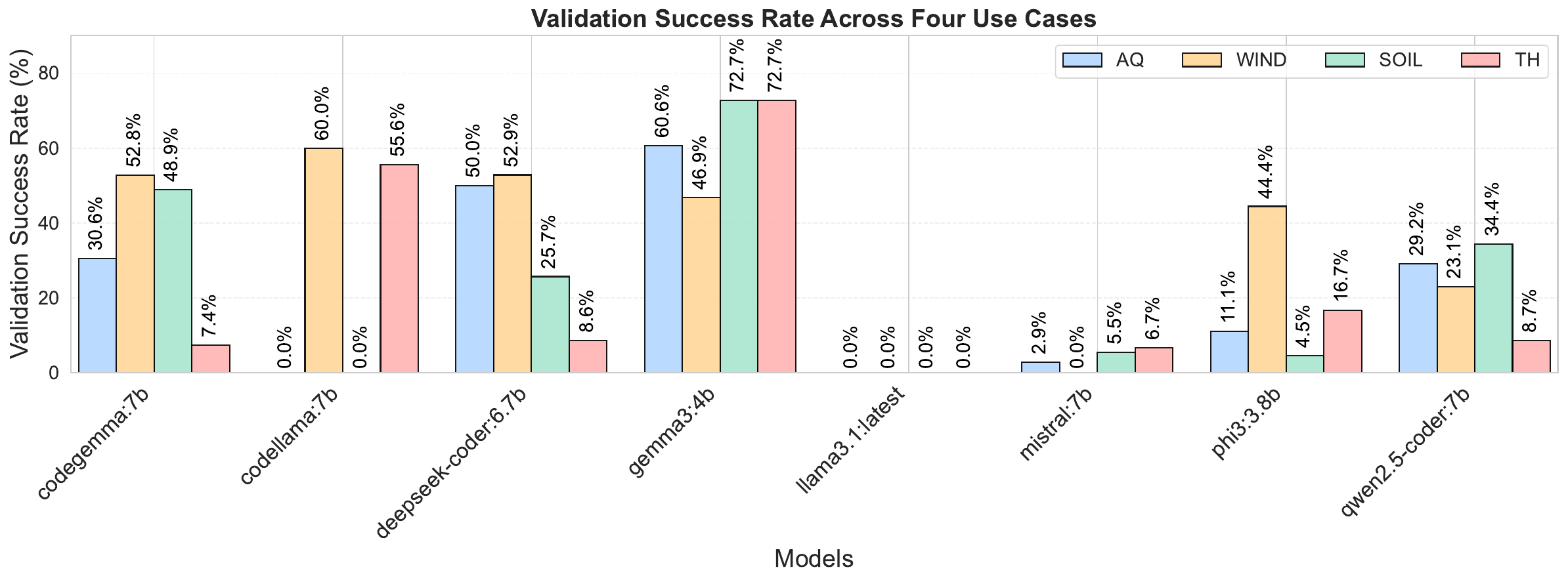}
    \caption{Validation success rate}
    \label{fig:val_result}
\end{figure}

Figure \ref{fig:val_result} illustrates the average validation success rate per run across four datasets. For each run, the success rate is calculated as the percentage of generated scripts that passed the validator (step 3, as shown in \ref{fig:sbs}) relative to the total number of scripts submitted to the validator. The key finding is that gemma3\_4B outperforms other models, despite having a lower parameter count, and achieves the highest validation success rate in three out of four cases. It reaches approximately 60.6\% in the AQ dataset, 72.2\% in the Soil dataset, 72.7\% in the TH dataset, and 46.9\% in the Wind dataset. This figure also displays the limitations of general-purpose models (less than 10B) in this framework. Models like mistral\_7b and phi\_3.8b exhibit the lowest validation success rate (below 12\%) in the AQ and Soil dataset, while llama3.1\_latest never reached Step 4. Surprisingly, codellama\_7b fails to generate code in Step 2 (code generation) for the AQ and Soil datasets, but shows promising results for Wind (60\% highest among all models) and for TH (55.6\% second-highest after gemma3\_4b). 

Models with a 0.0\% success rate across all use cases indicate that either no scripts reach the Scheduler (step 4) for execution (as in the case of llama3.1:lastest \& mistral:7b) or none are successfully executed. We also note that, despite deepseek-coder generating more scripts reaching Step 4 than the gemma3:4b model, the validation success rate of gemma3:4b is higher than deepseek-coder:6.7b and outperforms other models across many use cases while taking less latency in many of them. Even codellama:7b generates a few scripts that reached step 4, but its validation success rate is better than deepseek-coder while maintaining lower end-to-end latency as shown in Figure \ref{fig:latency}.

\subsubsection{Execution success rate}
\begin{figure}[htbp]
    \centering
    \includegraphics[width=0.95\linewidth]{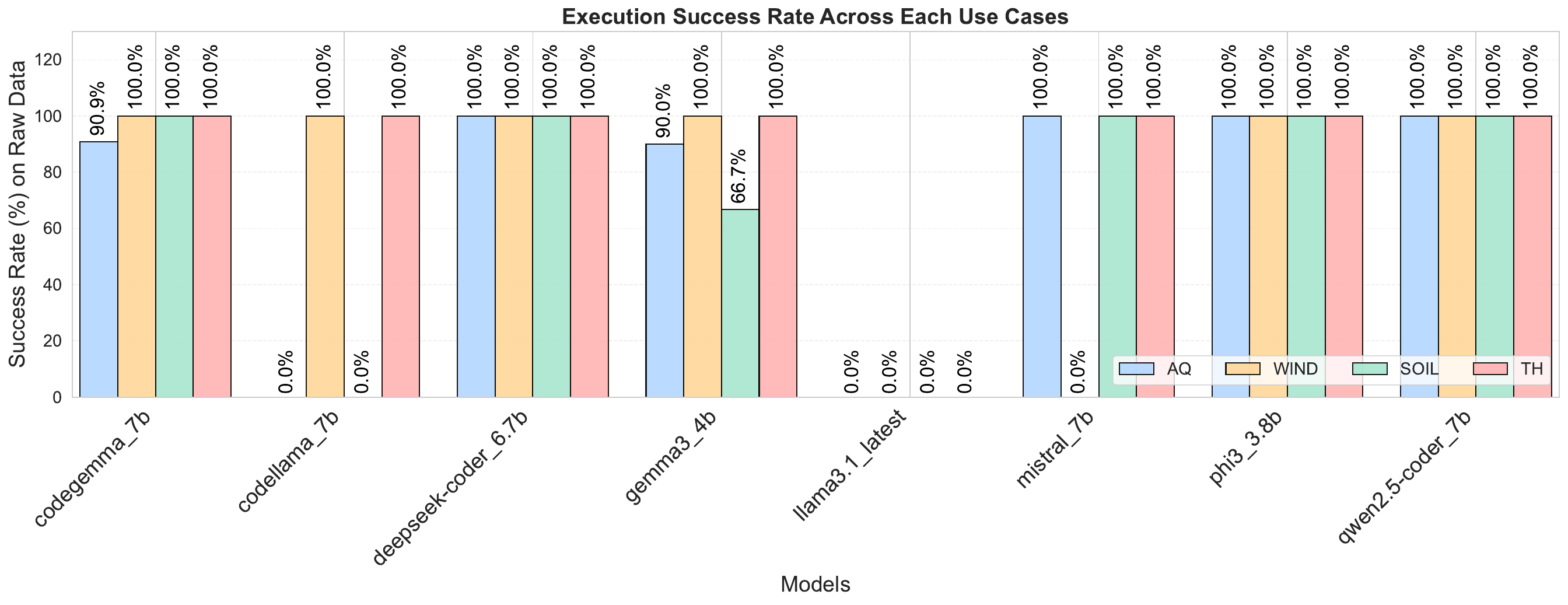}
    \caption{Execution success rate}
    \label{fig:step4_result}
\end{figure}

Figure \ref{fig:step4_result} presents the final execution success rate of different models across four use cases (AQ, WIND, SOIL, TH). The Execution pass rate is calculated as a percentage of the number of scripts that executed successfully by the scheduler on the raw data to the number of scripts that reached Step 4 (shown in Fig. \ref{fig:sbs}). The result shows that most of the models achieve 100\% success rate with the exception of codegemma\_7b for the AQ dataset (90.9\%) and gemma3\_4b for the AQ (90.0\%) \& SOIL (66.7\%) datasets. We also note that many models show 0\% validation success rates, including llama3.1-latest across all use cases, codellama:7b for the AQ \& SOIL dataset, and mistral:7b for the WIND dataset. For most models, the execution success rate is 100\%, indicating that our proposed framework eliminates most faulty scripts before Step 4, leaving only correct scripts to be stored and executed on the raw data.

\section{Future Work} \label{sec:fw}
In current framework, the scheduler executes tasks sequentially and edge devices work independently, which may limits efficiently. In the future, we plan to implement dynamic scheduling using LLMs and allowing edge devices to share knowledge and improve performance together. Finally, we will explore specialized models for specific tasks, for example using a general-purpose model for task generation and a code-based model for code generation.

\section{Conclusion}\label{sec:con}
This paper presented the LEI framework, an agentic architecture designed to enable autonomous, context-aware, and self-upgradable edge systems. LEI addresses the limitations of static, hard-coded edge intelligence by introducing a cognitive orchestration layer that enables cloud-hosted LLMs to dynamically interpret sample data, metadata, contextual information, and real-time resource constraints to generate optimized, task-specific programs for edge devices. By shifting from manually defined business logic to adaptive intelligence generation, LEI reduces developer intervention, enhances flexibility, and supports the continuous evolution of edge capabilities. The framework considers hardware heterogeneity and resource constraints, enabling efficient business logic tailored to device-specific conditions while preserving low latency and operational efficiency. LEI is implemented on small-scale experimental testbeds comprising Raspberry Pi devices (RPi 4B and RPi 5) and private cloud environments. The evaluation is conducted using real-world environmental monitoring use cases, including air quality, temperature–humidity, soil, and wind analysis. The framework is evaluated using metrics such as token generation rate (tokens per second), end-to-end latency, reliability (success rate of LLM-generated code), and overall system robustness.

\backmatter

\section*{Declarations}

\begin{itemize}
\item Funding: None
\item Conflict of interest/Competing interests: There is no conflict of interest between authors. 
\item Data availability : Data used in this paper is approriately cited within the paper. 
\item Code availability: Our implementations are publicly available in Github repository via link \url{https://github.com/chinmaya-dehury/LEI-LLM-assistedEI}
\end{itemize}

\noindent
If any of the sections are not relevant to your manuscript, please include the heading and write `Not applicable' for that section. 






\begin{appendices}

\section{Air Quality}\label{air_quality}
\subsection{Metadata of Air Quality} \label{app:metadata_AirQuality}

\begin{lstlisting}[
    breaklines=true, 
    basicstyle=\tiny\ttfamily, 
    literate=
        {µ}{{$\mu$}}1        % Replaces 'µ' with Greek math letter mu
        {³}{{$^3$}}1         % Replaces '³' with superscript 3
        {°}{{$^{\circ}$}}1,
    caption={Complete Metadata Structure}, 
    label={code:json_file}
]

{
  "data_source": "IoT air quality Sensors",
  "sampling_interval": "10-15 minutes",
  "data_format": "CSV",
  "columns": {
    "timestamp": {
      "type": "datetime",
      "description": "Time at which the reading was recorded (local time)."
    },
    "coord": [
      {
        "lon": "integer",
        "range":[0,180],
        "unit": "degrees",
        "description": "longitude of a given location."
      },
      {
        "lat": "integer",
        "range":[0,90],
        "unit": "degrees",
        "description": "latitude of a given location."
      }
    ],
    "components": [
      {
        "co": {
          "type": "float",
          "unit": "µg/m3",
          "description": "Carbon monoxide concentration."
        }
      },
      {
        "no": {
          "type": "float",
          "unit": "µg/m3",
          "description": "Nitric oxide concentration."
        }
      },
      {
        "no2": {
          "type": "float",
          "unit": "µg/m3",
          "description": "Nitrogen dioxide concentration."
        }
      },
      {
        "no3": {
          "type": "float",
          "unit": "µg/m3",
          "description": "Ozone concentration."
        }
      },
      {
        "so2": {
          "type": "float",
          "unit": "µg/m3",
          "description": "Sulphur dioxide concentration."
        }
      },
      {
        "pm2_5": {
          "type": "float",
          "unit": "µg/m3",
          "description": "PM2.5 concentration (fine particles <=2.5 µm)."
        }
      },
      {
        "pm10": {
          "type": "float",
          "unit": "µg/m3",
          "description": "PM10 concentration (particles <=10 µm)."
        }
      },
      {
        "nh3": {
          "type": "float",
          "unit": "µg/m3",
          "description": "Ammonia concentration."
        }
      },
      {
        "so2": {
          "type": "float",
          "unit": "µg/m3",
          "description": "Sulphur dioxide concentration."
        }
    },
      {
        "pm2_5": {
          "type": "float",
          "unit": "µg/m3",
          "description": "PM2.5 concentration (fine particles <=2.5 µm)."
        }
    },
      {
        "pm10": {
          "type": "float",
          "unit": "µg/m3",
          "description": "PM10 concentration (particles <=10 µm)."
        }
      },
      {
        "nh3": {
          "type": "float",
          "unit": "µg/m3",
          "description": "Ammonia concentration."
        }
      }
    ]    
  },
  "data_quality": {
    "missing_values": "none",
    "sensor_accuracy": "not available"
  },
  "device_specs": {
    "edge_device_model": "Raspberry Pi 4",
    "compute_capability": "Quad-core 1.5GHz, 4GB RAM"
  },
  "storage": "Local SQLite DB with periodic upload to cloud"
}
\end{lstlisting}

\subsection{Context of Air Quality} \label{app:context_AirQuality}
Below is the full content of the \texttt{context.txt} file of Air Quality provided to the LLM.

\begin{lstlisting}[
    % --- Text File Settings ---
    basicstyle=\small\ttfamily, 
    breaklines=true,             
    columns=fullflexible,        
    keepspaces=true,             
    caption={Complete Air Quality Context},
    label={code:context_txt}
]
This sample dataset represents the air quality of a specific location (or for a coordinate on the globe).
The IoT edge device collects current air pollution data in 10-15 minutes.
The purpose is to monitor the air quality and obtain the insights that are useful for citizens.
Example of some insights can be real-world health warnings, potential source, etc. 
Typical insights could include (but not limited to):
-> AQI level trends - overall air quality score (Good-Severe).
-> Main pollutant identification - which pollutant contributes most to poor air.
-> Peak pollution hours - when air quality worsens (morning/evening).
-> Source inference - traffic, industry, or agriculture dominance.
-> Seasonal variation - pollution increase in winter vs summer.
-> Health risk mapping - respiratory and cardiovascular hazard levels.

The LLM should generate small, efficient Python programs suitable for execution on the edge device.
Each program should focus on one insight at a time and be optimized for limited computational resources.
\end{lstlisting}




\end{appendices}

\bibliography{ref}

\end{document}